\documentclass[aps,prd,twocolumn,floats,floatfix,nofootinbib]{revtex4-1}
\usepackage{graphicx}
\usepackage{amssymb}
\usepackage{epstopdf}
\usepackage{hyperref}
\usepackage{multirow}
\usepackage{float}

\usepackage{amsmath}
\usepackage{bbold}
\usepackage{color}

\begin{document}

\title{Multi-component dark matter from a hidden gauged SU(3)}

\author{Alexandre Poulin}
\email{apoulin@physics.carleton.ca}

\author{Stephen Godfrey}
\email{godfrey@physics.carleton.ca}

\affiliation{Ottawa-Carleton Institute for Physics, Carleton University, 1125 Colonel By Drive, 
Ottawa, Ontario K1S 5B6, Canada}

\date{\today}                                  

\begin{abstract}
We study Dark Matter (DM) phenomenology with multiple DM species consisting of both
scalar and vector DM particles.  
More specifically, we study the Hidden Gauged SU(3) model of Arcadi {\it et al}.  
Before proceeding to the Hidden Gauged SU(3) model, we study the relic abundances of simplified
multi-species DM scenarios to gain some insights when multiple species and interactions are included.
In the Hidden Gauged SU(3) model,
because of the large parameter space, we restrict ourselves to 
three representative benchmark points, each with multiple DM species. 
The relic densities for the benchmark points 
were found using a program developed to solve the coupled Boltzmann equations for an 
arbitrary number of interacting DM species with two particles in the final state. For each case, 
we varied the mass of the DM particles and then found the value of the dark SU(3) gauge coupling 
that gives the correct relic density.  
We found that in some regions of the parameter space, the DM would be difficult to
observe in direct detection experiments while easier to observe in indirect detection experiments
and vice versa,
so that complementary measurements could help pinpoint the details of the Hidden Gauged SU(3) model.
Important to this, is that even for moderate changes in input parameter values,
the relative relic density of each species can change significantly resulting in large
changes in the observability of multi-species DM by direct or indirect detection. 
\end{abstract}

\maketitle

\section{Introduction}

The nature of Dark Matter (DM) is one of the biggest puzzles in particle physics.  
The evidence for DM in the Universe comes from a wide variety of astrophysical and 
cosmological observations and there has been considerable theoretical effort to 
understand its nature (for recent reviews, see for example Ref.~\cite{Gelmini:2015zpa,Bertone:2004pz,Feng:2010gw,Roszkowski:2017nbc,Arcadi:2017kky}).  
Up until now, much of the work has focused on simple DM sectors,
typically with only one or possibly two DM candidates.  These scenarios are increasingly constrained 
by experimental measurements.  Therefore, workers in the field are exploring ever more
complex DM scenarios including models with multiple DM species (see for example,
\cite{Arcadi:2016kmk,Gross:2015cwa,Dienes:2011ja,Baer:2011uz,Chialva:2012rq,KeithBrooks:2012,Zurek:2008qg,Esch:2014jpa,Elor:2015bho,Boddy:2014yra,DiFranzo:2016uzc,Bhattacharya:2016ysw,Ko:2016fcd,Feldman:2010wy,DEramo:2010keq,Aoki:2012ub,Dienes:2016vei,Dienes:2016kgc,Forestell:2016qhc,Karam:2016rsz,Bae:2014rfa,Hur:2007ur,SungCheon:2008ts,Dienes:2014via,Aoki:2014lha,Dienes:2011sa,Aoki:2016glu,Dienes:2017ylr,Dienes:2017zjq,Bhattacharya:2017fid,Ahmed:2017dbb,Arcadi:2017vis,Aoki:2018gjf}).   
In this context, there is a growing body of work studying details of ``freezeout'' for multiple interacting DM species,
e.g. \cite{Edsjo:1997bg,DEramo:2010keq,Baer:2011uz,Aoki:2012ub,Bae:2014rfa,Forestell:2016qhc,Bhattacharya:2016ysw,Karam:2016rsz,Dienes:2017zjq,DHT2018,Feldman:2010wy,Ahmed:2017dbb}. 
To expand on this theme, we developed a computer program capable of calculating the 
relic densities of an arbitrary number of interacting DM species and, in this work, use it to explore the behavior for several
scenarios and the implications for DM direct and indirect detection experiments.

As a specific example of a model with multiple DM species, we
re-examine the Hidden Gauged SU(3) model of
Arcadi {\it et al.} \cite{Arcadi:2016kmk}. This model has the potential to have numerous 
DM species depending on the choice of parameter values 
from the breaking of the new SU(3).
The paper of Arcadi {\it et al.} \cite{Arcadi:2016kmk} chose specific values of the parameters
that reduced the multi-species DM problem to effectively two DM species to enable them to implement
the model in the computer program MicrOMEGAs \cite{Belanger:2018ccd}.  It turns out that 
different values of the parameters 
 can lead to 
 much more complex and interesting scenarios with multiple DM species to explore.  
Thus, a main focus of this paper is to explore different values of the parameters 
of the  Hidden Gauged SU(3) model that lead to a more complex DM particle spectrum.  
We found that some scenarios with
multiple DM species give the correct relic density, but because the dominant
components have a small direct detection cross section, the DM will be difficult to observe in
direct detection experiments. However, in some cases when the DM is not observable by direct detection 
experiments, it could be observable in indirect detection measurements and vice versa.
These points were also made by Arcadi {\it et al.} \cite{Arcadi:2016kmk}
and by many others for other DM scenarios.  A small sampling of examples is given by 
Ref.~\cite{Gelmini:2015zpa,Bertone:2004pz,Feng:2010gw,Roszkowski:2017nbc,Arcadi:2017kky,KeithBrooks:2012,Dienes:2014via,Dienes:2017ylr}.

The Hidden Gauged SU(3) model has a very rich phenomenology with details of the phenomenology very
much dependent on the choice of parameter values. 
Some of our results convey a warning that approximations
can lead to erroneous conclusions.  
For example, in a scenario with multiple DM species, a moderate change in
the input parameter values can lead to major differences in the relative
relic densities of the stable DM species.  This would have a major impact on whether 
DM could be observed by direct or indirect detection experiments.
An important lesson from our study is that models with multiple species of DM particles
can lead to a very rich phenomenology and that one needs to be 
careful when studying relic densities in multi-species DM models.

The parameter space of the Hidden Gauged SU(3) model is large.  Integrating the coupled 
Boltzmann equations to obtain the relic density
for multiple DM species is computationally intensive
so that it is impractical to perform a complete scan of the parameter space.  Instead, we
choose a small number of representative benchmark points in the parameter space and examine
their phenomenology. For two of the benchmark points, there are 4 stable DM
species, while in the third, there are 7.  
These cases will be examined in detail in Section~\ref{sec:Discussion}.

Section II gives the details of the coupled Boltzmann equations we use
to calculate the relic abundance for multi-species DM scenarios \cite{DHT2018}.  This is the core ingredient 
of the paper needed to calculate the subsequent results for the Hidden Gauged SU(3) model.
Before proceeding to the Hidden Gauged SU(3) model,
we explore two simple multi-species DM scenarios in Section~\ref{sec:simple_scenario}.
In that section, we assume three or four scalar DM species with 
a range of masses and assumptions for the interactions between them.  We do so to gain
some insights before proceeding to the more complicated Hidden Gauged SU(3) model.  
In Section~\ref{sec:model},
we give the details of the Hidden Gauged SU(3) model.  In that section,
we start by writing down the Lagrangian, then compute the mass eigenstates for the scalar and 
vector sectors in terms of the Lagrangian parameters, the vacuum expectation values, and the new gauge
coupling.  We then write down expressions for the theoretical constraints on these parameters. 
In Section~\ref{sec:Discussion}, we explore the phenomenology of three benchmark 
points for the Hidden Gauged SU(3) model  that satisfy
the theoretical constraints and use the relic density to constrain the parameters of the model 
which are then used to calculate direct detection and indirect detection cross sections.
Finally, in Section~\ref{sec:conclusions}, we summarize our results.

\section{The Boltzmann Equations for Multi-species DM}
\label{sec:freezeout}

Our purpose is to study models of DM that include multiple DM species.
The first step in such a study is to calculate 
the relic density for a DM sector with an arbitrary number 
of interacting DM species.  
To calculate the DM relic density, we assume that the dark sector was in thermal
and chemical equilibrium at some early time which allows us to
use the coupled Boltzmann equations in the form written down 
by Dienes, Huang, and Thomas \cite{DHT2018}
which is a generalization of the Boltzmann equation given in Kolb and Turner
\cite{Kolb:1990vq} (see also Ref.~\cite{Gondolo:1990dk}).  
These themes have been explored previously 
by Edsjo and Gondolo \cite{Edsjo:1997bg}  and others, for example \cite{DEramo:2010keq,DiFranzo:2016uzc,Bae:2014rfa,Forestell:2016qhc,Karam:2016rsz,Ahmed:2017dbb,Griest:1990kh}.

In what follows, we simply reproduce the coupled Boltzmann equations as given by Dienes, Huang, and Thomas \cite{DHT2018} 
which we will use to calculate each of the individual relic densities and refer the interested reader to their work 
for details   \cite{DHT2018}.
We only include the details needed to reproduce our results including some
of the equations and definitions that we used from Kolb and Turner \cite{Kolb:1990vq} and refer the
interested reader to that monograph for more complete details.

We wish to calculate the relic density of each of the DM species, $\phi_i$, which is given by
\begin{align}
\Omega_i=\frac{\rho_i}{\rho_C}=\frac{8\pi Gm_i n_i}{3H^2},
\end{align}
where $m_i$, $n_i$, and $\rho_i=m_i n_i$ are the mass, number density,
and energy density of particle $\phi_i$, $\rho_C$ is the critical density, 
$G$ is Newton's constant,
and $H$ is the Hubble parameter. For species $\phi_i$ in kinetic equilibrium, the number density
is given by
\begin{equation}
n_i^{\rm eq} = {{g_i}\over{(2\pi)^3}}\int  f_i(\vec{p}_i) d^3p,
\end{equation}
where $g_i$ is the number of internal degrees of freedom and the phase space 
distribution, $f_i(\vec{p}_i)$, is given by
\begin{equation}
f_i(\vec{p}_i)= {{1}\over {e^{(E_i - \mu_i)/T} \pm 1}},
\end{equation}
where $T$ is the temperature of the photon bath,
 $\vec{p}_i$, $E_i$, and $\mu_i$ are the three-momentum, energy, and chemical potential 
associated with species $\phi_i$, and
the plus and minus signs are for fermionic and bosonic species respectively.
We are working in the comoving frame in which $\langle \vec{v} \rangle =0$ for all $\phi_i$. 
In a frame in which $\langle \vec{v} \rangle =0$,
 the $f_i(\vec{v})$ only depend on the magnitude of $\vec{v}$ and not its direction. 

In general, the Boltzmann equations  can include an arbitrary
number of species participating in reactions and decays \cite{DHT2018}.
We will restrict ourselves to simplified cases where:
the Boltzmann equations for all relevant species can be expressed in terms of number
densities, $n_i$, rather than the phase space distributions, $f_i$;  we assume $CP$ 
is conserved so that the matrix elements satisfy 
$|{\cal M}(ij\to kl)|^2 = |{\cal M}(kl \to ij)|^2$ for cross sections 
and $|{\cal M}(i\to jk)|^2 = |{\cal M}(jk \to i)|^2$ for 
decays; and that all visible sector particles and all dark sector 
particles are in thermal equilibrium due to rapid interactions among themselves until long 
after the lightest DM species have frozen out. 
To minimize the complexity of the problem,
we restrict ourselves to $1\to 2$ and $2\to 2$ type interactions.   
With these simplifications, the Boltzmann equations 
become \cite{DHT2018}: 
\begin{widetext}
\begin{eqnarray}\label{eqn:td}
{{dn_i}\over{dt}} &= & 
- 3H n_i - \sum_{a,b} \Gamma_{i\to a b } (n_i - n_i^{eq}) 
- \sum_{j\ne i} \sum_a \left[ { \Gamma_{i\to j a} \left( { n_i - n_j {{n_i^{eq}}\over{n_j^{eq}}} } \right)
- \Gamma_{j\to i a} \left( { n_j - n_i {{n_j^{eq}}\over{n_i^{eq}}} } \right) }\right] \\
& &-\sum_{j,k}\left[\frac{1}{2}(1+\delta_{jk})\Gamma_{i\to jk}\left(n_i-n_jn_k\frac{n_i^{eq}}{n_j^{eq}n_k^{eq}}\right)
- (1+\delta_{ik})\Gamma_{j\to ik}\left(n_j-n_in_k\frac{n_j^{eq}}{n_i^{eq}n_k^{eq}}\right)\right] \nonumber \\
& & - \sum_j \sum_{a,b} (1+\delta_{ij}) \langle \sigma_{ij\to ab} v \rangle (n_i n_j -n_i^{eq} n_j^{eq}) \nonumber \\
& & - \sum_j \sum_{a,b} \left[ { \langle \sigma_{ia\to jb} v \rangle 
\left( { n_i  - n_j {{n_i^{eq}}\over{ n_j^{eq}} } } \right)  n_a^{eq}
- \langle \sigma_{ja\to ib} v \rangle 
\left( { n_j  - n_i {{n_j^{eq}}\over{ n_i^{eq}} } } \right) n_a^{eq} } \right] \nonumber \\
& & - \sum_{j,k} \sum_a \left[ {  (1+\delta_{jk}) \langle \sigma_{ia\to jk} v \rangle 
\left( {n_i  - n_j n_k { {n_i^{eq}} \over{ n_j^{eq}  n_k^{eq}} } } \right)  n_a^{eq}
- (1+\delta_{ik} ) \langle \sigma_{ja\to ik} v \rangle 
\left( { n_j  - n_i n_k {{n_j^{eq}}\over{ n_i^{eq} n_k^{eq} } } }\right) n_a^{eq} } \right] \nonumber \\
& & - \sum_{j,k} \sum_a   (1+\delta_{ij} )  \langle \sigma_{ij\to ka } v \rangle
\left( { n_i n_j  - n_k {{n_i^{eq} n_j^{eq}}\over{ n_k^{eq}} } } \right)
  - \sum_{j,k,l} (1+\delta_{ij} )(1+\delta_{kl})
   \langle \sigma_{ij\to kl } v \rangle
\left( { n_i n_j  - n_k n_l {{n_i^{eq} n_j^{eq}}\over{ n_k^{eq} n_l^{eq} } } } \right),
  \nonumber 
\end{eqnarray}
\end{widetext}
where in Eq.~\ref{eqn:td}, the indices $i,j,k,l$ run over the dark-sector particles and $a,b$ run over visible sector particles.
The first term on the right-hand side is the dilution due to the expansion of the Universe and
$H$ is the Hubble parameter.  $\langle \sigma_{xy\to zw}v \rangle$, which will be given below,
represents the thermally averaged annihilation cross section of two initial state particles, $xy$,
to  two final state particles, $zw$, where $xyzw$ represent either visible or dark sector particles.  
Likewise, $ \Gamma_{x\to yz} $ represents the partial decay width of particle $x$ into particles $yz$.
Eq.~\ref{eqn:td} 
assumes that all particles are in thermal equilibrium. In Eq.~\ref{eqn:td}, the thermally averaged cross sections are given by:
\begin{equation}
\langle \sigma_{xy\to zw} v_{xy} \rangle \equiv {{g_x g_y}\over {n_x^{eq} n_y^{eq} }} 
\int \sigma_{xy\to zw} v_{xy} f_x^{eq} f_y^{eq} 
{{d^3p_x}\over{(2\pi)^3}}  {{d^3p_y}\over{(2\pi)^3}},
\label{eqn:sigtave}
\end{equation}
where $g_i$ is the number of internal degrees of freedom of particle $i$,
$\sigma_{xy\to zw}$ is the cross section for the process $\phi_x \phi_y \to \phi_z \phi_w$
and 
\begin{equation}
v_{xy} \equiv |v_x-v_y| = {{\sqrt{(p_x \cdot p_y)^2 - m_x^2 m_y^2}}\over{E_x E_y}}
\end{equation}
is the magnitude of the relative velocity of $\phi_x$ and $\phi_y$ in the comoving frame.  
In the absence of a degenerate Fermi species or a Bose condensate, the phase space distributions 
of the DM species reduce to the classical
Maxwell-Boltzmann distribution in the non-relativistic limit \cite{Kolb:1990vq}. 
For sufficiently rapid interactions, the particles approach kinetic and 
chemical equilibrium depending on the nature of the interaction.
This latter situation places constraints on the chemical potentials of the species involved which depend on the reaction. For example, the reaction $\phi_i+\phi_j\leftrightarrow \phi_k+\phi_l$ would imply the constraint $\mu_i+\mu_j=\mu_k+\mu_l$. For the visible particles, the chemical potential $\mu$ is small or zero which results in the chemical potential of the DM species being small or zero in the absence of an asymmetry. We will therefore treat all equilibrium chemical potentials as zero.
Under these conditions we approximate $f_i(v_i)$ by the equilibrium phase space distribution:
\begin{equation}
f_i^{eq}(v_i) \approx e^{-E_i(v_i)/T}
\end{equation}
where $v_i$ is the speed of the particle.
With these approximations, the equilibrium number density, $n_i^{eq}$, is given by 
\begin{align}
n_i^{eq} =& g_i \int {{d^3p_i}\over {(2\pi)^3}} e^{-E_i/T} = {{g_i m_i^2 T}\over {2\pi^2}} K_2(m_i/T),\\
\approx& \begin{cases}g_i \left(\frac{m_iT}{2\pi}\right)^{3/2}e^{-m_i/T} \mbox{ for } m_i\gg T,\\
g_i\frac{T^3}{\pi^2} \mbox{ for } T\gg m_i,
\end{cases}
\label{eqn:neq}
\end{align}
where $K_2(x)$ is the modified Bessel function of the second kind.

The standard approach for solving Eq.~\ref{eqn:td} is
to scale out the effect of the expansion
of the Universe by considering the evolution of the number density in a comoving volume.
This is done by using the entropy density, $s$, as a fiducial quantity by defining
 \cite{Kolb:1990vq}:
\begin{equation}
Y={n_i\over s}, \quad Y^{eq}= {n^{eq}_i \over s},
\end{equation}
where the entropy density in the comoving volume in any cosmological epoch is defined by
\begin{equation}
s \equiv {S\over V} = {{\rho +p}\over T} = {{2\pi^2}\over{45}} g_{*S}T^3,
\label{eqn:s}
\end{equation}
$T$ is the photon temperature and 
$\rho$ and $p$ are the total energy density and the pressure, respectively, 
expressed in terms of the photon temperature $T$ \cite{Kolb:1990vq}.
$g_{*S}$ represents the number of interacting degrees of freedom in the thermal 
bath which can be found by rewriting Eq.~\ref{eqn:s} as
\begin{equation}
g_{*S}= { {45}\over {2\pi^2 T^4}} (\rho +p). \label{eqn:g*}
\end{equation}
$Y_i$ is the actual number per comoving volume and $Y^{eq}_i$ is the equilibrium
number per comoving volume.  
In addition, we rescale the 
Hubble constant $H(T)=H(m)/x^2$ with $x=m/T$ where $m$ is any convenient mass scale \cite{Kolb:1990vq}
which we will take to be the mass of the heaviest stable DM particle. 
In the radiation dominated epoch, $H(m)$ is given by \cite{Kolb:1990vq}
\begin{align}
H(m) =\frac{2\pi^{3/2}}{\sqrt{45}}  {{ g_*^{1/2} m^2} \over {m_{Pl}  }},
\end{align} 
where $m_{Pl}$ is the Planck mass and $g_*$ 
represents the total number of effectively massless degrees of freedom and is given by 
\begin{equation}
g_* = {{30 \rho_R}\over{\pi^2 T^4}},
\end{equation}
where $\rho_R$ is the total energy density of all species in equilibrium.
Expressions for $\rho_R$, $g_{*S}$, and $g_*$ can be found in Kolb and Turner \cite{Kolb:1990vq}. 
In computing $g_*$ and $g_{*S}$ we used
the Fermi-Dirac or Bose-Einstein distributions as appropriate in the sum. For the region around
the QCD phase transition, the correct degrees of freedom  becomes ambiguous and requires
lattice QCD to compute. We use the values found in  \cite{Steigman} for this region.

With these substitutions, Eq.~\ref{eqn:td} is recast as:
\begin{widetext}
\begin{eqnarray}
\label{eqn:tdY}
{{dY_i}\over{dx}} &= & 
 - {{xs}\over{H(m)} } \left\{ {
  {1\over s}  \sum_{a,b} \Gamma_{i\to a b } (Y_i - Y_i^{eq}) 
+{1\over s} \sum_{j\ne i} \sum_a \left[ { \Gamma_{i\to j a} \left( { Y_i - Y_j {{Y_i^{eq}}\over{Y_j^{eq}}} } \right)
- \Gamma_{j\to i a} \left( { Y_j - Y_i {{Y_j^{eq}}\over{Y_i^{eq}}} } \right) }\right] } \right. \\
& &-{1\over s}\sum_{j,k}\left[\frac{1}{2}(1+\delta_{jk})\Gamma_{i\to jk}\left(Y_i-Y_jY_k\frac{Y_i^{eq}}{Y_j^{eq}Y_k^{eq}}\right)
- (1+\delta_{ik})\Gamma_{j\to ik}\left(Y_j-Y_iY_k\frac{Y_j^{eq}}{Y_i^{eq}Y_k^{eq}}\right)\right] \nonumber \\
& & + \sum_j \sum_{a,b} (1+\delta_{ij}) \langle \sigma_{ij\to ab} v \rangle (Y_i Y_j -Y_i^{eq} Y_j^{eq}) \nonumber \\
& & + \sum_j \sum_{a,b} \left[ { \langle \sigma_{ia\to jb} v \rangle 
\left( { Y_i  - Y_j {{Y_i^{eq}}\over{ Y_j^{eq}} } } \right)  Y_a^{eq}
- \langle \sigma_{ja\to ib} v \rangle 
\left( { Y_j  - Y_i {{Y_j^{eq}}\over{ Y_i^{eq}} } } \right) Y_a^{eq} } \right] \nonumber \\
& & + \sum_{j,k} \sum_a \left[ {  (1+\delta_{jk}) \langle \sigma_{ia\to jk} v \rangle 
\left( {Y_i  - Y_j Y_k { {Y_i^{eq}} \over{ Y_j^{eq}  Y_k^{eq}} } } \right)  Y_a^{eq}
- (1+\delta_{ik} ) \langle \sigma_{ja\to ik} v \rangle 
\left( { Y_j  - Y_i Y_k {{Y_j^{eq}}\over{ Y_i^{eq} Y_k^{eq} } } }\right) Y_a^{eq} } \right] \nonumber \\
& & + \sum_{j,k} \sum_a  (1+\delta_{ij} )  \langle \sigma_{ij\to ka } v \rangle
\left( { Y_i Y_j  - Y_k {{Y_i^{eq} Y_j^{eq}}\over{ Y_k^{eq}} } } \right)
+ \left.   \sum_{j,k,l} (1+\delta_{ij} )(1+\delta_{kl})
 {  \langle \sigma_{ij\to kl } v \rangle
\left( { Y_i Y_j  - Y_k Y_l {{Y_i^{eq} Y_j^{eq}}\over{ Y_k^{eq} Y_l^{eq} } } } \right)
} \right\} . \nonumber 
\end{eqnarray}
\end{widetext}

\section{Simple Multi-component DM Scenarios}
\label{sec:simple_scenario}

\begin{figure*}[h]
\begin{center}
\includegraphics[scale=0.47,keepaspectratio=true]{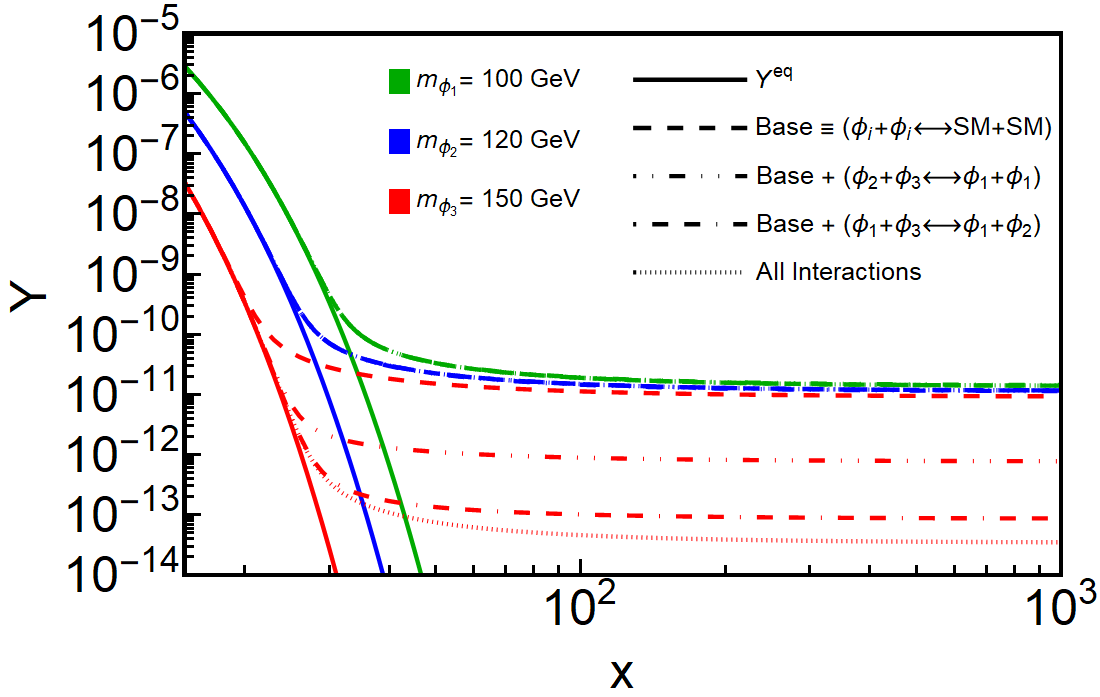} 
\end{center}
\vskip -0.1in
   \caption{ 
   DM number densities as a function of $x=m/T$ for three DM species. The solid curves represent the equilibrium
   density, the dashed curves represent the scenario with only self-annihilation to SM particles which 
   we label as the ``base'',
    the curves with both dots and dashes represent scenarios with the base interactions plus a single additional interaction as labeled in
    the plot, and the dotted curve represent the result including all the additional interactions of the previous curves. The green, blue, and red curves 
    represent the number density of $\phi_1$, $\phi_2$, and $\phi_3$ with masses of $100$ GeV, $120$ GeV, and $150$ GeV, respectively.
    Note that for the green and blue curves, the various curves sit very closely on top of each other.
}
\label{fig:mcfoa}
\end{figure*}

\begin{figure*}[h]
\begin{center}
\includegraphics[scale=0.47,keepaspectratio=true]{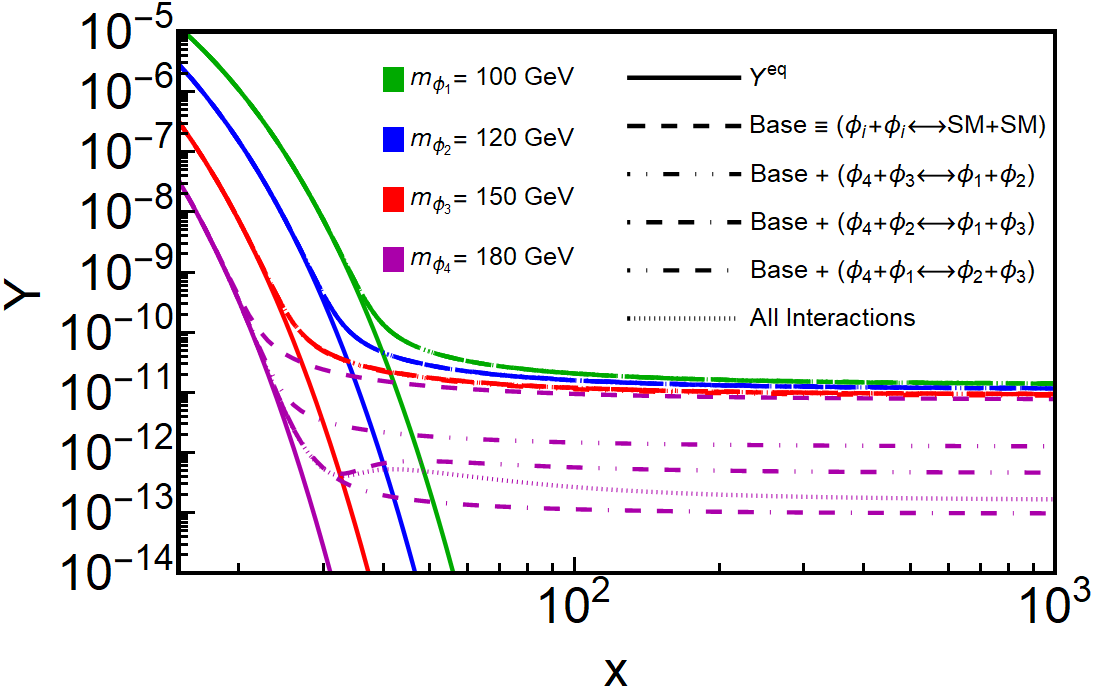} 
\end{center}
\vskip -0.1in
   \caption{ 
   DM number densities as a function of $x=m/T$ for four DM species. The line labelling is as in Fig.~\ref{fig:mcfoa}.
     The green, blue, red, and purple curves 
    represent the number density of $\phi_1$, $\phi_2$, $\phi_3$, and $\phi_4$ with masses of $100$ GeV, $120$ GeV, $150$ GeV, and $180$ GeV, respectively.
    Note that for the green, blue, and red curves, the various curves sit very closely on top of each other.
}
\label{fig:mcfob}
\end{figure*}

Before proceeding to the Hidden Gauged SU(3) DM model, we start by exploring the relic 
abundance of two simplified multi-species DM scenarios.  Models with multiple species 
and different types of interactions can be complicated.  Deconstructing how the different
terms in Eq.~ \ref{eqn:tdY} affect the number densities can give us insights into the
coupled Boltzmann equations and help us understand which terms are
important and which ones can be neglected.
Both of the scenarios initially consist of multiple DM species freezing out 
by only self-annihilating into Standard Model (SM) particles. We then add additional 
interactions between the DM species and observe the effects. Finally, we include all interactions that
would be present due to crossing symmetry to obtain the final result.
We assume that the thermally averaged cross sections for all interactions are s-wave, 
and thus independent of speed, and use the approximation $\langle \sigma v \rangle = 0.1$ pb 
for all the interactions. In a particular model,
these cross sections would not necessarily be equal, but this assumption is reasonable 
for the purpose of exploring the behavior of Eq. \ref{eqn:tdY} for simplified scenarios.
In all cases, we assume that the SM particles are in thermal equilibrium with the 
photon bath and treat them as relativistic with the appropriate degrees of freedom.  Plots
of the DM number per comoving volume versus $x=m/T$ 
are shown in Figures \ref{fig:mcfoa} and \ref{fig:mcfob} where we take $m$ to be the mass of the heaviest stable
DM particle. 

In all cases, the solid curves 
represent the equilibrium number densities, the dashed lines represent the scenarios with only self-annihilation to 
SM particles which we label as the ``base'', 
the lines with a combination of dashes and
dots include the base interactions and a single additional interaction, ignoring the contributions 
from crossing symmetry, and finally the dotted lines show the results when all the additional interactions are included. 
We label the species as $\phi_i$ with the index $i$ increasing with increasing species mass.

In Fig.~\ref{fig:mcfoa}, we consider three DM species, $\phi_1$ (green), $\phi_2$ (blue), and $\phi_3$ (red),
with masses of $100$ GeV, $120$ GeV, and $150$ GeV, respectively.  
We will start with the most general application of Eq.~\ref{eqn:tdY} that is applicable to this case and
examine each of the contributions in turn.
For this case, Eq.~\ref{eqn:tdY} reduces to the following set of three coupled equations:
\begin{align}
{{dY_1}\over{dx}} = &  - {{xs}\over{H(m)} } \Bigg\{ 2 \langle \sigma_{11\to SM+SM} v \rangle (Y_1^2 -(Y_1^{eq})^2) \label{eqn:tdYa1}\\
& -2 \langle \sigma_{23\to 11 } v \rangle
\left( { Y_2 Y_3  - Y_1^2 {{Y_2^{eq} Y_3^{eq}}\over{ (Y_1^{eq})^2 } } } \right)
\Bigg\} , \nonumber \\
{{dY_2}\over{dx}} = &  - {{xs}\over{H(m)} } \Bigg\{ 2 \langle \sigma_{22\to SM+SM} v \rangle (Y_2^2 -(Y_2^{eq})^2) \label{eqn:tdYa2}\\
& + \langle \sigma_{23\to 11 } v \rangle
\left( { Y_2 Y_3  - Y_1^2 {{Y_2^{eq} Y_3^{eq}}\over{ (Y_1^{eq})^2 } } } \right)\nonumber\\
& - \langle \sigma_{13\to 12 } v \rangle
\left( { Y_3  -  Y_2 {{ Y_3^{eq}}\over{ Y_2^{eq} } } } \right)Y_1\Bigg\} , \nonumber \\
{{dY_3}\over{dx}} = &  - {{xs}\over{H(m)} } \Bigg\{ 2 \langle \sigma_{33\to SM+SM} v \rangle (Y_3^2 -(Y_3^{eq})^2) \label{eqn:tdYa3}\\
& + \langle \sigma_{23\to 11 } v \rangle
\left( { Y_2 Y_3  - Y_1^2 {{Y_2^{eq} Y_3^{eq}}\over{ (Y_1^{eq})^2 } } } \right)\nonumber\\
& + \langle \sigma_{13\to 12 } v \rangle
\left( { Y_3  -  Y_2 {{ Y_3^{eq}}\over{ Y_2^{eq} } } } \right)Y_1\Bigg\} \nonumber.
\end{align}
From this set of equations we can examine which factors influence each of the DM species' number densities.

Starting with $\phi_1$,
the only additional term that can affect the $\phi_1$ number density is  $\phi_2+\phi_3\leftrightarrow\phi_1+\phi_1$
because $\phi_1$ only acts as a spectator in the $\phi_1+\phi_3\leftrightarrow\phi_1+\phi_2$ interaction 
leaving the $\phi_1$ number density unchanged. 
The interesting temperature is when $\phi_1$ freezes out when the actual $\phi_1$ number density departs from the equilibrium
number density.  This occurs when $Y_1=cY_1^{eq}$ for  $c \gtrsim  1$  \cite{Kolb:1990vq}.
Because $\phi_1$ is lighter than the other DM species, $\phi_2$ and $\phi_3$ 
have already departed from chemical equilibrium so that $Y_2\gg Y_2^{eq}$ and $Y_3\gg Y_3^{eq}$. This gives:
\begin{align} \label{eq:fig1-c1}
Y_1^2 -(Y_1^{eq})^2\approx& (1-c^{-2})Y_1^2,\\
{ Y_2 Y_3  - Y_1^2 {{Y_2^{eq} Y_3^{eq}}\over{ (Y_1^{eq})^2 } } }\approx& Y_2Y_3\nonumber.
\end{align}
Since $Y_1\gg Y_2$ and $Y_1\gg Y_3$ at the temperature where $Y_1=cY_1^{eq}$, 
 the second line in Eq.~\ref{eq:fig1-c1} is negligible 
and can safely be ignored during a numerical scan.
This explains why the two additional terms in Fig.~\ref{fig:mcfoa} have
no significant effect on the $\phi_1$ relic density.

We next examine the temperature region around where  $\phi_2$ freezes out; $Y_2=cY_2^{eq}$.
In this case,  $Y_1\approx Y_1^{eq}$ and $Y_3\gg Y_3^{eq}$. This leads to:
\begin{align} \label{eq:fig1-c2}
Y_2^2 -(Y_2^{eq})^2\approx& (1-c^{-2})Y_2^2,\\
{ Y_2 Y_3  - Y_1^2 {{Y_2^{eq} Y_3^{eq}}\over{ (Y_1^{eq})^2 } } }\approx& Y_2Y_3,\nonumber\\
\left({ Y_3  -  Y_2 {{ Y_3^{eq}}\over{ Y_2^{eq} } } }\right) Y_1\approx& Y_1Y_3.\nonumber
\end{align}
Since $Y_2^2\gg Y_2Y_3$, the second line in Eq.~\ref{eq:fig1-c1} will be negligible.
The third line could potentially 
be important since $Y_1\gg Y_2 \gg Y_3$.  However it turns out that for our choice 
of masses the $\phi_1+\phi_3\leftrightarrow\phi_1+\phi_2$
interaction causes 
$Y_2^2\gg Y_1Y_3$ so this term also has a neglible contribution  compared to the first
line in Eq.~\ref{eq:fig1-c2}.  The net result is 
that the interactions added to the base case do not significantly alter the $\phi_2$
relic abundance.

Finally, we consider the temperature where $\phi_3$ freezes out;
 $Y_3=cY_3^{eq}$. In this case, $Y_1 \approx Y_1^{eq}$ and $Y_2 \approx Y_2^{eq}$. This gives:
\begin{align} \label{eq:fig1-c3}
Y_3^2 -(Y_3^{eq})^2\approx& (1-c^{-2})Y_3^2,\\
{ Y_2 Y_3  - Y_1^2 {{Y_2^{eq} Y_3^{eq}}\over{ (Y_1^{eq})^2 } } }\approx& (1-c^{-1})Y_2Y_3,\nonumber\\
\left({ Y_3  -  Y_2 {{ Y_3^{eq}}\over{ Y_2^{eq} } } }\right) Y_1\approx& (1-c^{-1})Y_1Y_3.\nonumber
\end{align}
In this case the largest reaction rate is for 
$\phi_1+\phi_3\leftrightarrow \phi_1+\phi_2$, corresponding to the third line of Eq.~\ref{eq:fig1-c3},
 followed by $\phi_2+\phi_3\leftrightarrow\phi_1+\phi_1$, corresponding to the second line of  Eq.~\ref{eq:fig1-c3},
because of the hierarchy of the number densities, $Y_1\gg Y_2\gg Y_3$. 
This is what we see in  Fig.~\ref{fig:mcfoa} where the curves with the additional interactions 
stay in thermal equilibrium longer, resulting in a smaller relic density for $\phi_3$.  
So for the case of annihilation of $\phi_3$ with $\phi_1$,
there is a small amount of $\phi_3$ which sees a large amount of $\phi_1$ so that the
interaction $\phi_1+\phi_3\leftrightarrow \phi_1+\phi_2$ depletes $\phi_3$ significantly but
barely affects the $\phi_1$ density which is what is seen in Fig.~\ref{fig:mcfoa}.  Because $\phi_2$
has a smaller relic density than $\phi_1$,  it doesn't deplete $\phi_3$ as much, which is also observed
in Fig.~\ref{fig:mcfoa}.

In  Fig.~\ref{fig:mcfob} we repeat the exercise  
for  four DM species, $\phi_1$ (green), $\phi_2$ (blue), $\phi_3$ (red), and $\phi_4$ (purple), 
with masses of $100$ GeV, $120$ GeV, $150$ GeV, and $180$ GeV,  respectively.  
For this case Eq.~\ref{eqn:tdY} leads to a set of four coupled equations. 
$\phi_1$, $\phi_2$, and $\phi_3$ behave similarly to the previous case so to avoid repetition 
we will focus on the behavior of the $\phi_4$ number density.
The differential  equation for $Y_4$ is given by:
\begin{align}
{{dY_4}\over{dx}} = &  - {{xs}\over{H(m)} } \Bigg\{ 2 \langle \sigma_{44\to SM+SM} v \rangle (Y_4^2 -(Y_4^{eq})^2) \label{eqn:tdYb1}\\
& + \langle \sigma_{34\to 12 } v \rangle
\left( { Y_3 Y_4  - Y_1Y_2 {{Y_3^{eq} Y_4^{eq}}\over{ Y_1^{eq}Y_2^{eq} } } } \right)\nonumber  \\
& + \langle \sigma_{24\to 13 } v \rangle
\left( { Y_2 Y_4  - Y_1Y_3 {{Y_2^{eq} Y_4^{eq}}\over{ Y_1^{eq}Y_3^{eq} } } } \right) \nonumber  \\
& + \langle \sigma_{14\to 23 } v \rangle
\left( { Y_1 Y_4  - Y_2Y_3 {{Y_1^{eq} Y_4^{eq}}\over{ Y_2^{eq}Y_3^{eq} } } } \right)
\Bigg\} . \nonumber 
\end{align}

To obtain some insights into the behavior of the $\phi_4$ number density  we start by
considering the temperature where $\phi_4$ freezes out, $Y_4=cY_4^{eq}$ with $c\gtrsim 1$,
and assume that in this region $Y_1\approx Y_1^{eq}$, $Y_2\approx Y_2^{eq}$, and $Y_3\approx Y_3^{eq}$:
\begin{align}
Y_4^2 -(Y_4^{eq})^2\approx& (1-c^{-2})Y_4^2,\\
{ Y_3 Y_4  - Y_1Y_2 {{Y_3^{eq} Y_4^{eq}}\over{ Y_1^{eq}Y_2^{eq} } } } \approx& (1-c^{-1})Y_3Y_4,\nonumber\\
{ Y_2 Y_4  - Y_1Y_3 {{Y_2^{eq} Y_4^{eq}}\over{ Y_1^{eq}Y_3^{eq} } } } \approx& (1-c^{-1})Y_2Y_4,\nonumber\\
{ Y_1 Y_4  - Y_2Y_3 {{Y_1^{eq} Y_4^{eq}}\over{ Y_2^{eq}Y_3^{eq} } } } \approx& (1-c^{-1})Y_1Y_4.\nonumber
\end{align}
These expressions  can explain why the $\phi_4+\phi_3\leftrightarrow \phi_1+\phi_2$ (dot-dot-dashed) 
curve sits above the $\phi_4+\phi_2\leftrightarrow \phi_1+\phi_3$ (dot-dashed-dashed)
and $\phi_4+\phi_1\leftrightarrow \phi_2+\phi_3$ (dot-dot-dashed-dashed) curves, but fails to 
explain other details of this plot.  As pointed out above, the number density of a DM species depends
on its mass  so that there is less of $\phi_3$ to annihilate with than
 $\phi_1$ and $\phi_2$ so that $\phi_4$ is depleted less by its iteractions with $\phi_3$ than
with the other DM species. 

Next, we consider the behavior of the $\phi_4$ number density for the curves
when the $\phi_4+\phi_2\leftrightarrow\phi_1+\phi_3$ and $\phi_4+\phi_1\leftrightarrow\phi_2+\phi_3$  interactions
are present in the temperature region around $\phi_3$ and $\phi_4$ freezeout 
 where
 $Y_4=c_1Y_4^{eq}$ and $Y_3=c_2Y_3^{eq}$ with $c_1 \simeq c_2 \gtrsim 1$:
\begin{align} \label{eqn:fig2:b}
Y_4^2 -(Y_4^{eq})^2\approx& (1-c^{-2}_1)Y_4^2,\\
{ Y_3 Y_4  - Y_1Y_2 {{Y_3^{eq} Y_4^{eq}}\over{ Y_1^{eq}Y_2^{eq} } } } \approx& (1-c_1^{-1}c_2^{-1})Y_3Y_4,\nonumber\\
{ Y_2 Y_4  - Y_1Y_3 {{Y_2^{eq} Y_4^{eq}}\over{ Y_1^{eq}Y_3^{eq} } } } \approx& (1-c_2c_1^{-1})Y_2Y_4,\nonumber\\
{ Y_1 Y_4  - Y_2Y_3 {{Y_1^{eq} Y_4^{eq}}\over{ Y_2^{eq}Y_3^{eq} } } } \approx& (1-c_2c_1^{-1})Y_1Y_4.\nonumber
\end{align}
In this temperature region the  factors, $(1-c_2 c_1^{-1})$, in the third and fourth equations in Eq.~\ref{eqn:fig2:b},
are small so the $\phi_4$ number density is dominated by the first equation
in Eq.~\ref{eqn:fig2:b}. The 2nd equation is not relevant here as it applies to the
$\phi_4+\phi_3\leftrightarrow\phi_1+\phi_2$ case which we are not considering here.  
The result is that the curves 
for $\phi_2+\phi_4\leftrightarrow\phi_1+\phi_3$ and $\phi_1+\phi_4\leftrightarrow\phi_2+\phi_3$ 
have similar values of $Y_4$ shortly after $\phi_4$ departs from the equilibrium curve. 

Lastly, we examine the region where $\phi_2$ freezes out corresponding to the temperature
where  $Y_2=cY_2^{eq}$ with $c\gtrsim 1$  as before. 
We assume  that $Y_1\approx Y_1^{eq}$,  $Y_3\gg Y_3^{eq}$ and $Y_4\gg Y_4^{eq}$ so that
\begin{align}
Y_4^2 -(Y_4^{eq})^2\approx& (1-c^{-2})Y_4^2,\\
{ Y_3 Y_4  - Y_1Y_2 {{Y_3^{eq} Y_4^{eq}}\over{ Y_1^{eq}Y_2^{eq} } } } \approx& Y_3Y_4,\nonumber\\
{ Y_2 Y_4  - Y_1Y_3 {{Y_2^{eq} Y_4^{eq}}\over{ Y_1^{eq}Y_3^{eq} } } } \approx& Y_2\left(Y_4-c^{-1}Y_3\frac{Y_4^{eq}}{Y_3^{eq}}\right),\nonumber\\
{ Y_1 Y_4  - Y_2Y_3 {{Y_1^{eq} Y_4^{eq}}\over{ Y_2^{eq}Y_3^{eq} } } } \approx& Y_1\left(Y_4-cY_3\frac{Y_4^{eq}}{Y_3^{eq}}\right).\nonumber
\end{align}
It turns out that  
 $Y_4$ is numerically comparable to $Y_3\frac{Y_4^{eq}}{Y_3^{eq}}$ at the temperature range we are considering.
As the temperature decreases, $c$, which is a measure of the difference between $Y$ and $Y^{eq}$, increases,
so that $\left(Y_4-cY_3\frac{Y_4^{eq}}{Y_3^{eq}}\right)$ will become negative and the reverse reaction will dominate. 
At the same time,  $\left(Y_4-c^{-1}Y_3\frac{Y_4^{eq}}{Y_3^{eq}}\right)$  remains positive and continues 
to deplete $\phi_4$.  This is why in Fig.~\ref{fig:mcfob}, the 
$\phi_4+\phi_1\leftrightarrow\phi_2+\phi_3$ (dot-dot-dashed-dashed) curve increases
 while the $\phi_4+\phi_2\leftrightarrow\phi_1+\phi_3$ (dot-dashed-dashed) curve decreases.
The dotted line shows the result when all processes are included.
It can be seen that the 
reverse reaction is important up until $\phi_1$ starts to freeze out, at which point the 
 $\phi_1+\phi_4\leftrightarrow\phi_2+\phi_3$ interaction becomes dominant and starts depleting $\phi_4$.

These simple examples  show that even for a simple multi-species DM sector the
behavior of the relic abundance can be quite complicated and the resulting 
relic abundances can vary substantially depending on the details of 
 the underlying model. It is not always a priori obvious how the details of the model will work out
 and one needs to understand how each of the interactions contribute to the final result.
  All things being equal, heavier DM species will be more affected
 by their interactions with lighter DM species than vice versa.   The
 interactions between multi-species SM can  lead to   experimental 
consequences.  Say, for example, that a model has many DM species but only one interacts with
a portal to the visible universe.  One could imagine a scenario where the model could 
account for the measured DM relic density, but the component that interacts with the visible
universe has a very small relic density, making it unlikely that it could be observed 
in direct or indirect detection experiments.  

In the next sections, we will explore in detail a specific multi-species DM model,
the Hidden Gauged SU(3) model. Because 
the parameter space is large, we will restrict ourselves to several representative benchmark points with interesting phenomenology.


\section{The Hidden Gauged SU(3) Model}
\label{sec:model}

In the remainder of this paper, we study the phenomenology of the Hidden Gauged SU(3) model
 of Arcadi {\it et al.} \cite{Arcadi:2016kmk} which consists of spin-1 and spin-0 states.  
In their paper, they examined two representative limiting cases that made their numerical 
analysis tractable.  We extend their studies to consider different points in the parameter 
space that leads to more complex DM scenarios.  
We mention that others have also studied hidden SU(3) models \cite{Karam:2016rsz,Ko:2016fcd,Arcadi:2017vis}.

\subsection{Details of the Model}

For completeness and clarity, we start by reproducing the details of the 
Hidden Gauged SU(3) model of Arcadi {\it et al.} \cite{Arcadi:2016kmk}.
The  model consists of a gauged SU(3) which is fully broken by two complex scalar 
triplets, $\Phi_1$ and $\Phi_2$, so that all the new gauge bosons acquire a mass. 
These new scalars are not charged under the SM gauge groups and can only interact with the SM
through the SM Higgs doublet. To simplify the Lagrangian and insure 
additional stable states, a $\mathbb{Z}_2$ symmetry is also imposed such that
\begin{align}\label{eqn:z2}
\Phi_1\rightarrow&-\Phi_1,\notag\\
\Phi_2\rightarrow&\Phi_2,
\end{align}
which has the effect of only including even powers of the scalar triplets in the Lagrangian.
Following Ref.~\cite{Arcadi:2016kmk}, and imposing this additional $\mathbb{Z}_2$ symmetry,
we write the Lagrangian as:
\begin{align}
\mathcal{L}=\mathcal{L}_{\rm SM}+\mathcal{L}_{\rm portal}+\mathcal{L}_{\rm hidden},
\end{align}
where
\begin{align}
-\mathcal{L}_{\rm SM}\supset& V_{\rm SM}=m_H^2|H|^2+\frac{\lambda_H}{2}|H|^4,\\
-\mathcal{L}_{\rm portal}=& V_{\rm portal}=\lambda_{H11}|H|^2|\Phi_1|^2+\lambda_{H22}|H|^2|\Phi_2|^2,\\
-\mathcal{L}_{\rm hidden}=&-\frac{1}{4}G_{\mu\nu}^a G^{\mu\nu a}+|D_\mu \Phi_1|^2+|D_\mu \Phi_2|^2-V_{\rm hidden},\\
V_{\rm hidden} =& m_{11}^2|\Phi_1|^2+m_{22}^2|\Phi_2|^2+\frac{\lambda_1}{2}|\Phi_1|^4+\frac{\lambda_2}{2}|\Phi_2|^4\\
+\lambda_3 & |\Phi_1|^2|\Phi_2|^2
+\lambda_4|\Phi_1^\dag\Phi_2|^2+\frac{\lambda_5}{2}\left[(\Phi_1^\dag\Phi_2)^2+h.c\right].
\end{align}
Here, $G_{\mu\nu}^a=\partial_\mu A_\nu^a -\partial_\nu A_\mu^a +\tilde{g}f^{abc}A^b_\mu A^c_\nu$ is the 
field strength tensor for the hidden SU(3) where $\tilde{g}$ is the gauge coupling and $f^{abc}$ are 
the SU(3) structure constants. The covariant derivative is given by $D_\mu=\partial_\mu -i\tilde{g}A^a_\mu t^a$ 
where $t_a$ are the Gell-Mann matrices (see for example Ref.~\cite{Patrignani:2016xqp} where 
the $t_a = {1\over 2}\lambda_a$).

In general, the scalar doublet and triplets will have the form:
\begin{align}
&H=\frac{1}{\sqrt{2}}\begin{pmatrix}
\phi^H_{1,r}+i \phi^H_{1,i}\\
v+\phi^H_{2,r}+i \phi^H_{2,i}\\
\end{pmatrix},\notag\\
&\Phi_1=\frac{1}{\sqrt{2}}\begin{pmatrix}
\phi_{1,r}^1+i\phi_{1,i}^1\\
\phi_{2,r}^1+i\phi_{2,i}^1\\
v_1+\phi_{3,r}^1+i\phi_{3,i}^1
\end{pmatrix},\  \notag\\
&\Phi_2=\frac{1}{\sqrt{2}}\begin{pmatrix}
\phi_{1,r}^2+i\phi_{1,i}^2\\
v_2+\phi_{2,r}^2+i\phi_{2,i}^2\\
v_3+\phi_{3,r}^2+i\phi_{3,i}^2
\end{pmatrix},\label{eq:phiGeneral}
\end{align}
where the $v_i$ are the vacuum expectation values (vevs), 
the superscript indicates to which of the $\Phi_i$ the field belongs, and the subscripts indicate 
the position in the triplet and whether it is the real or imaginary component. When all three new 
vevs are taken to be non-zero, we have some freedom in defining 
the Goldstone bosons of the hidden SU(3). 
One convenient way to define them is
\begin{align}
G^1=&\phi^2_{1,i},\ G^2=\phi^2_{1,r},\ G^3=\phi^2_{2,i},\ G^4=\phi^1_{1,i},\ G^5=\phi^1_{1,r},\notag\\
G^6=&\frac{v_1\phi^1_{2,i}+v_2\phi^2_{3,i}}{\sqrt{v_1^2+v_2^2}},\ G^7=\frac{v_1\phi^1_{2,r}+v_3\phi^2_{2,r}+v_2\phi^2_{3,r}}{\sqrt{v_1^2+v_2^2+v_3^2}},\  \notag\\
G^8=&\frac{v_2v_3\phi^1_{2,i}+(v_1^2+v_2^2)\phi^1_{3,i} +v_1v_3\phi^2_{3,i}}{\sqrt{(v_1^2+v_2^2)(v_1^2+v_2^2+v_3^2)}},
\end{align}
which leaves the physical states
\begin{align}
\varphi_1&=\frac{v_2\phi^1_{2,i}-v_3\phi^1_{3,i} +v_1\phi^2_{3,i}}{\sqrt{v_1^2+v_2^2+v_3^2}},\\
\varphi_2&=\frac{v_3\phi^1_{2,r}+v_1\phi^2_{2,r}}{\sqrt{v_1^2+v_3^2}},\\
\varphi_3&=\frac{v_1v_2\phi^1_{2,r}+v_2v_3\phi^2_{2,r}+(v_1^2+v_3^2)\phi^2_{3,r}}{\sqrt{(v_1^2+v_2^2)(v_1^2+v_2^2+v_3^2)}},\\
\varphi_4&=\phi^1_{3,r}.
\end{align}

We will see in section \ref{sec:alt_minima} that in order to have a consistent vacuum state, we must have
$v_1v_3(\lambda_4+\lambda_5)=0$. Since we require $v_1\neq 0$ to fully break the SU(3), we will study the
cases where we take $v_3=0$ and allow  $\lambda_4\neq-\lambda_5$.  
For completeness, we give the details of the $\lambda_4=-\lambda_5$ case in the appendix for 
the interested reader who may wish to study this possibility.

In the $v_3=0$ case, the resulting Goldstone bosons are given by
\begin{align}
G^1=&\phi^2_{1,i},\ G^2=\phi^2_{1,r},\ G^3=\phi^2_{2,i},\ G^4=\phi^1_{1,i},\ G^5=\phi^1_{1,r},\ \notag\\
G^8=&\phi^1_{3,i},\ G^6=s_\beta\phi^1_{2,i}+c_\beta\phi^2_{3,i},\ G^7=s_\beta\phi^1_{2,r}-c_\beta\phi^2_{3,r},
\end{align}
where we define $s_\beta=\sin \beta=\frac{v_1}{\sqrt{v_1^2+v_2^2}}$ 
and $c_\beta=\cos \beta=\frac{v_2}{\sqrt{v_1^2+v_2^2}}$. 
This leads to the convenient definition for the physical states given by
\begin{align}
\varphi_1=&\phi^1_{3,r},\ \varphi_2=\phi^2_{2,r},\ \varphi_3=c_\beta \phi^1_{2,r}+s_\beta \phi^2_{3,r},\ \nonumber \\
\varphi_4=&c_\beta \phi^1_{2,i}-s_\beta \phi^2_{3,i},
\end{align}
and the resulting form of the triplets in the unitary gauge is given by
\begin{align}
\Phi_1=\frac{1}{\sqrt{2}}\begin{pmatrix}
0\\
c_\beta(\varphi_3+i\varphi_4)\\
v_1+\varphi_1
\end{pmatrix},\ 
\Phi_2=\frac{1}{\sqrt{2}}\begin{pmatrix}
0  \\
v_2+\varphi_2\\
s_\beta(\varphi_3-i\varphi_4)
\end{pmatrix}.
\end{align}
Spontaneous symmetry breaking in the SM happens in the usual way 
so that the Higgs doublet in the unitary gauge is given by:
\begin{align}
H=\frac{1}{\sqrt{2}}\begin{pmatrix}
0\\
v+h
\end{pmatrix},
\end{align}
where $v$ is the SM vev. 

In the unitary gauge, we write down the scalar mass matrix in the form $\mathcal{L}\supset \frac{1}{2}\Phi^T m^2_{\rm scalar}\Phi$ where $\Phi^T=(h, \varphi_1, \varphi_2, \varphi_3, \varphi_4)$ are the scalars in the model and where 
\begin{widetext}
\begin{align}
m^2_{\rm scalar}=\begin{pmatrix}
\lambda_H v^2 & \lambda_{H11}v v_1& \lambda_{H22} v v_2 & 0 & 0\\
\lambda_{H11}v v_1& \lambda_1 v_1^2 &\lambda_3 v_1 v_2 & 0 & 0\\
\lambda_{H22} v v_2 &\lambda_3 v_1 v_2 & \lambda_2 v_2^2 & 0 & 0\\
0&0&0& \frac{(\lambda_4+\lambda_5)}{2}(v_1^2+v_2^2)&0\\
0&0&0&0& \frac{(\lambda_4-\lambda_5)}{2}(v_1^2+v_2^2)
\end{pmatrix}.
\end{align}
\end{widetext}

We see that because the second row of $\Phi_1$ and the third row of $\Phi_2$ do not have a vev, 
$\varphi_3$ and $\varphi_4$ cannot mix with any of the other scalars.
In addition, because we do not have CP-violation in the scalar sector,  $\varphi_3$ and $\varphi_4$ cannot mix with each other so both are mass eigenstates. 
To differentiate the mass states from the gauge states, we relabel the mass eigenstates as $\varphi_3=\mathcal{H}$ 
and $\varphi_4=\chi$. We label the remaining mass eigenstates as $h_1$, $h_2$, and $h_3$ and parameterize 
the mixing between them using three angles $\theta_1$, $\theta_2$, and $\theta_3$ as follows:
\begin{equation}
\begin{pmatrix}
m_{h_1}^2 &0 &0\\
0& m_{h_2}^2 &0 \\
0&0& m_{h_3}^2
\end{pmatrix} =U^T\begin{pmatrix}
\lambda_H v^2 & \lambda_{H11}v v_1& \lambda_{H22} v v_2 \\
\lambda_{H11}v v_1& \lambda_1 v_1^2 &\lambda_3 v_1 v_2 \\
\lambda_{H22} v v_2 &\lambda_3 v_1 v_2 & \lambda_2 v_2^2 
\end{pmatrix}U,
\end{equation}
where
\begin{align}
&U=Y_1X_2Y_3, \ 
Y_1=\begin{pmatrix}
c_1 & 0 & -s_1\\
0 & 1 & 0 \\
s_1 & 0 &c_1
\end{pmatrix},\ 
X_2=\begin{pmatrix}
1 & 0 & 0\\
0 & c_2 & -s_2 \\
0 & s_2 &c_2
\end{pmatrix},\ \notag\\
&Y_3=\begin{pmatrix}
c_3 & 0 & -s_3\\
0 & 1 & 0 \\
s_3 & 0 &c_3
\end{pmatrix}.
\end{align}
Here, $m_{h_i}$ is the mass of $h_i$, $s_i=\sin\theta_i$, $c_i=\cos\theta_i$, for $i=1,2,3$. 
The advantage of this parameterization is that, in the limit of small $\theta_2$, we 
can define an effective mixing angle $\theta=\theta_1+\theta_3$ which mixes $h$ and $\varphi_2$. 
This is convenient since the mixing between the SM Higgs and the dark 
scalars is constrained to be small to be consistent with precision electroweak 
measurements.  Ref.~\cite{Falkowski:2015iwa,Arcadi:2016kmk} finds that $\sin\theta \lesssim 0.3-0.4$ depending
on the mass of the second scalar. For all our benchmark points, we will take $\sin \theta =0.1$.

In the vector boson sector, there is only mixing between $A^3$ and $A^8$ through the following mass matrix
\begin{align}
\mathcal{L}\supset\frac{\tilde{g}}{8}\begin{pmatrix}
A^3_\mu & A^8_\mu
\end{pmatrix}
\begin{pmatrix}
v_2^2 & -\frac{v_2^2}{\sqrt{3}}\\
-\frac{v_2^2}{\sqrt{3}} & \frac{1}{3}(4v_1^2+v_2^2)
\end{pmatrix}
\begin{pmatrix}
A^3_\mu \\ A^8_\mu
\end{pmatrix}.
\end{align}
We define the mass eigenstates as:
\begin{align}
A_\mu^{3'}=c_\alpha A_\mu^3+A_\mu^8 s_\alpha,\\
A_\mu^{8'}=-s_\alpha A_\mu^3+A_\mu^8 c_\alpha,
\end{align}
where $c_\alpha=\cos \alpha$, $s_\alpha=\sin\alpha$, and 
\begin{align}
\alpha=\begin{cases}
\frac{1}{2}\arctan\left(\frac{\sqrt{3}v_2^2}{2v_1^2-v_2^2}\right) & \mbox{for }  v_2^2\leq 2v_1^2\\
\frac{1}{2}\arctan\left(\frac{\sqrt{3}v_2^2}{2v_1^2-v_2^2}\right)+\frac{\pi}{2} &\mbox{for }  v_2^2> 2v_1^2
\end{cases}.
\end{align}
We note that $\alpha\in \left(0,\frac{\pi}{3}\right)$ and that $t_\alpha=\tan\alpha>0$. 
We label all the mass eigenstates with a prime to distinguish between the mass and gauge
eigenstates.  The resulting vector boson masses are given by
\begin{align}
m_{A^{1'}}&=m_{A^{2'}}=\frac{\tilde{g}}{2}v_2,\ 
m_{A^{4'}}=m_{A^{5'}}=\frac{\tilde{g}}{2}v_1,\ \notag\\
m_{A^{6'}}&=m_{A^{7'}}=\frac{\tilde{g}}{2}\sqrt{v_1^2+v_2^2}\notag\\
m_{A^{3'}}&=\frac{\tilde{g} v_2}{2}\left(1-\frac{t_\alpha}{\sqrt{3}}\right)^{1/2},\  
m_{A^{8'}}=\frac{\tilde{g} v_1}{2}\left(1-\frac{t_\alpha}{\sqrt{3}}\right)^{-1/2}.
\end{align}
We can also write $m_{A^{3'}}$ and $m_{A^{8'}}$ explicitly in terms of the vevs as
\begin{align}
m_{A^{3'}}^2=\frac{\tilde{g}^2}{6}\left(v_1^2+v_2^2-\sqrt{v_1^4-v_1^2v_2^2+v_2^4}\right),\\
m_{A^{8'}}^2=\frac{\tilde{g}^2}{6}\left(v_1^2+v_2^2+\sqrt{v_1^4-v_1^2v_2^2+v_2^4}\right).
\end{align}
From this form, we can see that $A^{3'}_\mu$ is the lightest vector boson 
while $A^{8'}_\mu$ is the heaviest, or one of the heaviest. When $v_1=v_2$,
$A^{8'}_\mu$ becomes degenerate with $A^{6'}_\mu$ and $A^{7'}_\mu$.

This model has a global $\mathbb{Z}_2\times \mathbb{Z}_2'\times \mathbb{Z}_2''$ symmetry 
remaining from the breaking of the 
SU(3) with $\mathbb{Z}_2''$ being an additional $\mathbb{Z}_2$ to the model studied in Ref.~\cite{Arcadi:2016kmk}.

The $\mathbb{Z}_2$ symmetry is a subgroup of two  
remaining U(1) symmetries,  one for the vectors and one for the scalars. The U(1) for the vectors is characterized by the matrix
\begin{align}
U=e^{i\delta_1/3}\mbox{diag}(e^{-i\delta_1},1,1).
\end{align}
Under the transformation $A^a_\mu t^a\rightarrow UA^a_\mu t^a U^\dag$, setting 
$\delta_1=\pi$ takes $A^{1,2,4,5}_\mu\rightarrow-A^{1,2,4,5}_\mu$ and does 
not change the other vector fields. The U(1) for the scalars is described by taking 
$\Phi_{1,2}\rightarrow e^{i\delta_2}\Phi_{1,2}$. The $\mathbb{Z}_2$ subgroup corresponds to choosing $\delta_1=\pi$ and 
$\delta_2=\pi$. The $\mathbb{Z}_2'$ symmetry is associated with complex 
conjugation which is an outer automorphism of SU(3). The $\mathbb{Z}_2''$ symmetry is one
that does not appear in Ref.~\cite{Arcadi:2016kmk}. This symmetry is a result of having added the extra $\mathbb{Z}_2$
in equation \ref{eqn:z2}. This added symmetry results in the scalar potential being invariant under rephrasing  
$\phi_3$ and $\phi_4$. Independently, a U(1) subgroup of SU(3) characterized by the matrix
\begin{align}
U'=e^{i\delta_3/3}\mbox{diag}(1,e^{-i\delta_3},1)
\end{align}
leaves the the field strength tensor invariant. In order for the scalar kinetic terms to be invariant, we must make
a specific choice for the rephasing of $\phi_3$ and $\phi_4$, as well as for $\delta_3$. This results in the following
transformations:
\begin{align}
\phi_{3,4}\rightarrow& -\phi_{3,4}\notag\\
A^{1,2,6,7}_\mu\rightarrow& -A^{1,2,6,7}_\mu
\end{align}
We summarize the charges of the physical states under the $\mathbb{Z}_2\times\mathbb{Z}_2'\times\mathbb{Z}_2''$ 
symmetries in Table \ref{table:z2z2z2charges}. A minus sign in the table represents multiplying the field by $-1$ under the symmetry.

\begin{table}
\caption{Charges of the physical states under the $\mathbb{Z}_2\times\mathbb{Z}_2'\times\mathbb{Z}_2''$ symmetry.}
\begin{tabular}{ |c|c|c| }
\hline
Physical states &  $\mathbb{Z}_2\times\mathbb{Z}_2'\times\mathbb{Z}_2''$ \\
\hline\hline
$h_1$, $h_2$, $h_3$ & $(+,+,+)$\\\hline
$\mathcal{H}$, $A^{7'}_\mu$ & $(+,+,-)$\\\hline
$A^{3'}_\mu$, $A^{8'}_\mu$& $(+,-,+)$\\\hline
$A^{5'}_\mu$& $(-,+,+)$\\\hline
$A^{4'}_\mu$& $(-,-,+)$\\\hline
$A^{2'}_\mu$& $(-,+,-)$\\\hline
$\chi$,  $A^{6'}_\mu$ & $(+,-,-)$\\\hline
$A^{1'}_\mu$& $(-,-,-)$\\\hline
\end{tabular}
\label{table:z2z2z2charges}
\end{table}

From these charges, we can identify the expected DM species in this model. 
From the charges in Table \ref{table:z2z2z2charges}, we expect a minimum of $3$ stable DM species
and a maximum of $7$ DM species. Because 
the mass degeneracies, $m_{A^{1'}}=m_{A^{2'}}$ and $m_{A^{4'}}=m_{A^{5'}}$, 
put restrictions on which decays are kinematically allowed, 
the actual minimum number of stable DM species is $4$. 

\subsection{Theoretical Constraints on Lagrangian Parameters}
\label{sec:theory_contraints}

The Hidden Gauged SU(3) model has numerous parameters.  Much of the parameter space can be excluded
by theoretical consistency of the scalar potential and by experimental measurements.  In the 
following subsections, we lay out the details of the theoretical constraints.

\subsubsection{Unitarity}
\label{sec:unitarity_constraints}

The scalar couplings in the potential can be bounded by perturbative unitarity 
of the 2 $\rightarrow$ 2 scalar field and vector boson scattering amplitudes.

The partial wave amplitudes, $a_J$, are related 
to the matrix element of the process, $\mathcal{M}$, by:
\begin{equation}
	\mathcal{M}=16\pi\sum_J (2J+1)a_J P_J(\cos\theta),
\end{equation}
where $J$ is the (orbital) angular momentum and $P_J(\cos \theta)$ are the Legendre polynomials. 
Perturbative unitarity requires that the zeroth partial wave amplitude, $a_0$, 
satisfies $|a_0| \leq 1$ or $|{\rm Re} \, a_0| \leq \frac{1}{2}$. 
Because the 2 $\rightarrow$ 2 scalar field scattering amplitudes are real at tree level, 
we adopt the second, more stringent constraint. 
We will use this to constrain the magnitudes of the scalar quartic couplings. 

We work in the high energy limit where only four-point diagrams contribute 
to  $2\rightarrow 2$ scalar scattering since 
all diagrams involving scalar propagators are suppressed by the square of the collision energy. 
Thus, the dimensionful couplings $m_H^2$, $m_{11}$, and $m_{22}$ are not 
constrained directly by perturbative unitarity. In the high energy limit, it is valid to use the 
Goldstone bosons as the physical degrees of freedom instead of the longitudinally polarized vector 
bosons. We neglect scattering processes involving transversely polarized gauge 
 bosons and fermions. 

Under these conditions, only the zeroth partial wave amplitude contributes to $\mathcal{M}$ resulting in 
the constraint $|{\rm Re} \, a_0| < \frac{1}{2}$ being equivalent to $|\mathcal{M}| < 8\pi$. 
To obtain a minimal list of constraints, we must ensure that all the eigenvalues of the coupled-channel scattering 
matrix $\mathcal{M}$, which includes each possible combination of two scalar fields in the initial and 
final states, satisfy this constraint. We include a symmetry factor of $1/\sqrt{2}$ for each pair of identical particles in the 
initial or final states. 

\begin{table}[t]
\caption{Charges of the scalar gauge eigenstates under the nine global $\mathbb{Z}_2$ symmetries of the quartic part of the scalar potential.}
\begin{tabular}{ |c|c|c|c|c|c|c|c|c|c| }
\hline
Fields &  \multicolumn{9}{|c|}{$\mathbb{Z}_2$ Charges} \\
\hline\hline
$\phi^H_{1,r}$& $-$& $+$& $+$& $+$& $+$& $+$& $+$& $+$& $+$\\\hline
$\phi^H_{1,i}$& $+$& $-$& $+$& $+$& $+$& $+$& $+$& $+$& $-$\\\hline
$\phi^H_{2,r}$& $+$& $+$& $-$& $+$& $+$& $+$& $+$& $+$& $+$\\\hline
$\phi^H_{2,i}$& $+$& $+$& $+$& $-$& $+$& $+$& $+$& $+$& $-$\\\hline
$\phi^1_{1,r}$& $+$& $+$& $+$& $+$& $-$& $+$& $+$& $-$& $+$\\\hline
$\phi^1_{1,i}$& $+$& $+$& $+$& $+$& $-$& $+$& $+$& $-$& $-$\\\hline
$\phi^1_{2,r}$& $+$& $+$& $+$& $+$& $+$& $-$& $+$& $-$& $+$\\\hline
$\phi^1_{2,i}$& $+$& $+$& $+$& $+$& $+$& $-$& $+$& $-$& $-$\\\hline
$\phi^1_{3,r}$& $+$& $+$& $+$& $+$& $+$& $+$& $-$& $-$& $+$\\\hline
$\phi^1_{3,i}$& $+$& $+$& $+$& $+$& $+$& $+$& $-$& $-$& $-$\\\hline
$\phi^2_{1,r}$& $+$& $+$& $+$& $+$& $-$& $+$& $+$& $+$& $+$\\\hline
$\phi^2_{1,i}$& $+$& $+$& $+$& $+$& $-$& $+$& $+$& $+$& $-$\\\hline
$\phi^2_{2,r}$& $+$& $+$& $+$& $+$& $+$& $-$& $+$& $+$& $+$\\\hline
$\phi^2_{2,i}$& $+$& $+$& $+$& $+$& $+$& $-$& $+$& $+$& $-$\\\hline
$\phi^2_{3,r}$& $+$& $+$& $+$& $+$& $+$& $+$& $-$& $+$& $+$\\\hline
$\phi^2_{3,i}$& $+$& $+$& $+$& $+$& $+$& $+$& $-$& $+$& $-$\\\hline
\end{tabular}
\label{table:z2FieldCharge}
\end{table}

We use the fields in the form used in equation \ref{eq:phiGeneral}.
This leads to a large $120\times 120$ dimension scattering matrix.  However, 
the part of the scalar potential that leads to the quartic interactions
 obeys nine $\mathbb{Z}_2$ discrete symmetries which 
bring the matrix $\mathcal{M}$ into block diagonal form which greatly simplifies the problem.
In a $2\rightarrow 2$ 
scattering process, the initial and final states must have the same charges under these $\mathbb{Z}_2$ 
symmetries. 
The charges we assign for the scalar fields that leave the
potential invariant under these symmetries  are given in Table~\ref{table:z2FieldCharge}. 
We can understand the origins of these charges by examining the various terms in the potential. 
The first four symmetries arise because the doublet, $H$, only appears 
as $|H|^2=\frac{1}{2}((\phi^H_{1,r})^2+(\phi^H_{1,I})^2+(\phi^H_{2,r})^2+(\phi^H_{2,I})^2)$ so 
changing the sign of any one of the fields does not change the term. The fifth through seventh 
symmetries can be understood by looking at the quadratic terms involving the triplets which appear in the potential, 
namely $|\Phi_1|^2$, $|\Phi_2|^2$, and $\Phi_1^\dag\Phi_2$. From this we see that there are
never any terms which multiply fields
from different scalar triplet rows. The eighth symmetry is 
related to the extra $\mathbb{Z}_2$ symmetry given in Eq.~\ref{eqn:z2} 
that we imposed on the potential. Finally, the last symmetry comes 
from taking the complex conjugate of the potential.
The resulting constraints are given by:
\begin{align}
|\lambda_H|\leq 8\pi,\\
|\lambda_{H11}|\leq 8\pi,\\
|\lambda_{H22}|\leq 8\pi,\\
|\lambda_3-\lambda_4|\leq 8\pi,\\
|\lambda_3+\lambda_4|\leq 8\pi,\\
|\lambda_3-\lambda_5|\leq 8\pi,\\
|\lambda_3+\lambda_5|\leq 8\pi,\\
|\lambda_3+3\lambda_4-4\lambda_5|\leq 8\pi,\\
|\lambda_3+3\lambda_4+4\lambda_5|\leq 8\pi,\\
|\lambda_1+\lambda_2+\sqrt{(\lambda_1-\lambda_2)^2+4\lambda_4^2}|\leq 4\pi,\\
|\lambda_1+\lambda_2-\sqrt{(\lambda_1-\lambda_2)^2+4\lambda_4^2}|\leq 4\pi,\\
|\lambda_1+\lambda_2+\sqrt{(\lambda_1-\lambda_2)^2+4\lambda_5^2}|\leq 4\pi,\\
|\lambda_1+\lambda_2-\sqrt{(\lambda_1-\lambda_2)^2+4\lambda_5^2}|\leq 4\pi,\\
|\mbox{Roots}(P(z))|\leq 8\pi, \label{eqn:polynomial}
\end{align}
where
\begin{align}
P(z)&=z^3-2z^2(4\lambda_1+4\lambda_2+3\lambda_H) \notag\\&-4z\Big((3\lambda_3+\lambda_4)^2-12(\lambda_1+\lambda_2)\lambda_H\notag\\&
-16\lambda_1\lambda_2+6\lambda_{H11}^2+6\lambda_{H22}^2\Big)\notag\\
&-24\Big(-(3\lambda_3+\lambda_4)^2\lambda_H+16\lambda_1\lambda_2\lambda_H-8\lambda_2\lambda_{H11}^2)\notag\\&-8\lambda_1\lambda_{H22}^2+4(3\lambda_3+\lambda_4)\lambda_{H11}\lambda_{H22}\Big).
\end{align}
This last polynomial comes from finding the eigenvalues of the 3 by 3 matrix:
\begin{align*}
P(z)=\det\begin{pmatrix}
8\lambda_1-z & 2(3\lambda_3+\lambda_4) & \sqrt{24}\lambda_{H11}\\
2(3\lambda_3+\lambda_4) & 8\lambda_2-z & \sqrt{24}\lambda_{H22}\\
\sqrt{24}\lambda_{H11}& \sqrt{24}\lambda_{H22} & 6\lambda_H-z
\end{pmatrix}.
\end{align*}
The bounds from Eq.~\ref{eqn:polynomial} 
can be translated into the following bounds using the same technique used in Ref.~\cite{Campbell:2016}:
\begin{widetext}
\begin{align*}
4\pi\geq&|2\lambda_1+2\lambda_2-\sqrt{4(\lambda_1-\lambda_2)^2+(3\lambda_3+\lambda_4)^2}|,\\
4\pi\geq&|2\lambda_1+2\lambda_2+\sqrt{4(\lambda_1-\lambda_2)^2+(3\lambda_3+\lambda_4)^2}|,\\
\lambda_H<&\frac{4\pi}{3}+\frac{4\left((3\lambda_1+\lambda_4)\lambda_{H11}\lambda_{H22}+2\lambda_{H11}^2(\pi-\lambda_2)+2\lambda_{H22}^2(\pi-\lambda_1)\right)}{(3\lambda_1+\lambda_4)^2-16(\pi-\lambda_1)(\pi-\lambda_2)},\\
\lambda_H>&-\frac{4\pi}{3}+\frac{4\left((3\lambda_1+\lambda_4)\lambda_{H11}\lambda_{H22}+2\lambda_{H11}^2(-\pi-\lambda_2)+2\lambda_{H22}^2(-\pi-\lambda_1)\right)}{(3\lambda_1+\lambda_4)^2-16(-\pi-\lambda_1)(-\pi-\lambda_2)}.
\end{align*}
\end{widetext}

\subsubsection{Scalar Potential Bounded from below}
\label{sec:bfb_constraints}

We next study the constraints on the scalar couplings by requiring that the potential be bounded from below. 
The constraints that must be satisfied at tree level for the scalar potential to be bounded from below can 
be determined by only considering the quartic terms of the potential because these dominate at large field values. 
We follow the approach of Ref.~\cite{Campbell:2016} and \cite{Arhrib:2011}. First, we make the following definitions:
\begin{align}
r=&\sqrt{|H|^2+|\Phi_1|^2+|\Phi_2|^2},\\
r^2\cos^2\gamma_1=&|H|^2,\\
r^2\cos^2\gamma_2\sin^2\gamma_1=&|\Phi_1|^2,\\
r^2\sin^2\gamma_2\sin^2\gamma_1=&|\Phi_2|^2,
\end{align}
\begin{align}
\zeta=&\frac{|\Phi_1^\dag\Phi_2|^2}{|\Phi_1|^2|\Phi_2|^2},\\
\omega=&\frac{(\Phi_1^\dag\Phi_2)^2+(\Phi_2^\dag\Phi_1)^2}{2|\Phi_1|^2|\Phi_2|^2}.
\end{align}
The parameters $\zeta$ and $\omega$ are bounded by
\begin{align}
\zeta\in& [0,1],\\
\omega\in& [-1,1].
\end{align}
Making these substitutions, we can write the quartic part of the potential as
\begin{align}
V_4=\frac{r^4}{(1+\tan^2\gamma_2)^2(1+\tan^2\gamma_1)^2}\mathbf{x}^T A\mathbf{y},\label{eq:V4}
\end{align}
where
\begin{align}
\mathbf{x}=\begin{pmatrix}
1\\
\tan^2\gamma_2\\
\tan^4\gamma_2
\end{pmatrix}, \ \mathbf{y}=\begin{pmatrix}
1\\
\tan^2\gamma_1\\
\tan^4\gamma_1
\end{pmatrix},
\end{align}
and
\begin{align}
A=\frac{1}{2}\begin{pmatrix}
\lambda_H& \lambda_{H11} &\lambda_1\\
2\lambda_H &\lambda_{H11}+\lambda_{H22}& 2(\lambda_3+\zeta\lambda_4+\omega\lambda_5)\\
\lambda_H&\lambda_{H22} &\lambda_2
\end{pmatrix}.
\end{align}
The fraction in Eq.~\ref{eq:V4} is always positive and grows
with the overall field excursion $r$. The $\mathbf{x}^T A\mathbf{y}$ term in
Eq. \ref{eq:V4} can be positive or negative; we require it to
be positive to ensure that the potential is bounded from
below. This term can be expressed as a biquadratic in $\tan\gamma_2$ with coefficients being 
other biquadratics in $\tan\gamma_1$.
A biquadratic of the form $a + bz^2 + cz^4$ will be positive
for all values of $z$ if the following conditions are satisfied:
\begin{align}
a>0, \ c>0, \mbox{ and } b+2\sqrt{ac}>0.
\end{align}
This provides us with the following constraints
\begin{align}
\lambda_H>&0,\\
\lambda_1>&0,\\
\lambda_2>&0,\\
\lambda_{H11}+2\sqrt{\lambda_1\lambda_H}>&0,\\
\lambda_{H22}+2\sqrt{\lambda_2\lambda_H}>&0,\\
\lambda_3+\zeta\lambda_4+\omega\lambda_5+\sqrt{\lambda_1\lambda_2}>&0.
\end{align}
This last bound can be split up into the different cases
\begin{align}
\lambda_3+\sqrt{\lambda_1\lambda_2}>\begin{cases}
\lambda_5 & \mbox{ for } \lambda_4>0,\ \lambda_5>0,\\
-\lambda_4+\lambda_5& \mbox{ for } \lambda_4<0,\ \lambda_5>0,\\
-\lambda_5& \mbox{ for } \lambda_4>0,\ \lambda_5<0,\\
-\lambda_4-\lambda_5& \mbox{ for } \lambda_4<0,\ \lambda_5<0.
\end{cases}
\end{align}

\begin{center}
\begin{table*}[t]
\caption{Input parameters for the benchmark points. $m_\chi$, $m_{\mathcal{H}}$, and $m_{{A^1}'}$ were allowed to 
vary for benchmark
point A, B, and C, respectively, along with the gauge coupling which was varied to obtain the correct relic density.
In all cases, we take the mixing in the scalar sector to only
be between $h_1$ and $h_3$ with a mixing angle of $\sin\theta=0.1$. We also set $m_{h_1}$ 
equal to the SM Higgs mass.}
\begin{tabular}{ |c|c|c|c| }
\hline
Parameters &  Scenarios A & Scenarios B & Scenarios C \\
\hline\hline
$m_{{A^1}'}$ & $300$ GeV & $300$ GeV & $200-500$ GeV  \\\hline
$m_{h_2}$ & $2500$ GeV & $2500$ GeV & $600$ GeV \\\hline
$m_{h_3}$ & $650$ GeV & $650$ GeV & $225$ GeV \\\hline
$m_{\mathcal{H}}$ & $1000$ GeV & $400 - 600$ GeV & $250$ GeV\\\hline
$m_{\chi}$ & $50 - 200$ GeV & $1000$ GeV & $250$ GeV \\\hline
$v_1/v_2$ & 10 & 10 & 1.2  \\\hline
DM & $\chi$, ${A^1}'$, ${A^2}'$, ${A^3}'$& $\mathcal{H}$, ${A^1}'$, ${A^2}'$, ${A^3}'$ & $\mathcal{H}$, $\chi$, ${A^1}'$, ${A^2}'$,\\
 &  & &  ${A^3}'$, ${A^4}'$, ${A^5}'$\\\hline
\end{tabular}
\label{table:benchmark}
\end{table*}
\end{center}

\subsubsection{Avoiding alternative minima}
\label{sec:alt_minima}

In this section, we discuss the structure of the vacuum of the model to confirm that the
choice of vevs corresponds to the absolute minimum of the potential. 
First, we write the potential in terms of the vevs:
\begin{align}
V(v,v_1&,v_2,v_3)= \frac{1}{2}\left(m_{H}^2v^2+m_{11}^2v_1^2+m_{22}^2(v_2^2+v_3^2)\right)\notag\\&+\frac{1}{8}\left(\lambda_Hv^4+\lambda_1v_1^4+\lambda_2\left(v_2^4+v_3^4\right)\right)\notag\\&+\frac{1}{4}\left(\lambda_{H11}v_1^2+\lambda_{H22}\left(v_2^2+v_3^2\right)\right)v^2\notag\\&+\frac{1}{4}\left(\lambda_3v_1^2v_2^2+\lambda_2v_2^2v_3^2+\left(\lambda_3+\lambda_4+\lambda_5\right)v_1^2v_3^2\right).
\end{align}
Minimizing the potential leads to the following constraints:
\begin{widetext}
\begin{align}
\frac{\partial V}{\partial v}=&0=v\left(2m_H^2+\lambda_H v^2 +\lambda_{H11}v_1^2+\lambda_{H22}(v_2^2+v_3^2)\right), \label{eqn:constraint1} \\
\frac{\partial V}{\partial v_1}=&0=v_1\left(2m_{11}^2+\lambda_{H11} v^2 +\lambda_1v_1^2+\lambda_3v_2^2+\left(\lambda_3+\lambda_4+\lambda_5\right)v_3^2\right),\\
\frac{\partial V}{\partial v_2}=&0=v_2\left(2m_{22}^2+\lambda_{H22} v^2 +\lambda_3v_1^2+\lambda_2(v_2^2+v_3^2)\right),\\
\frac{\partial V}{\partial v_3}=&0=v_3\left(2m_{22}^2+\lambda_{H22} v^2 +(\lambda_3+\lambda_4+\lambda_5)v_1^2+\lambda_2(v_2^2+v_3^2)\right). \label{eqn:constraint2}
\end{align}
\end{widetext}
Note that
$v_2\frac{\partial V}{\partial v_3}=v_3\frac{\partial V}{\partial v_2}+(\lambda_4+\lambda_5)v_1^2v_3$. 
From this, we conclude that to have a consistent vacuum configuration requires that 
$(\lambda_4+\lambda_5)v_1^2v_3=0$. Since we need $v_1\neq 0$ to fully break the hidden
SU(3) and setting $\lambda_4=-\lambda_5$ is an unmotivated tuned choice, we will assume that $v_3=0$. 
Given a collection of input parameters, we wish to ensure that we are in the deepest minima. From the inputs,
we can obtain values for the scalar couplings, the vevs, and $\tilde{g}$, but not $m_H^2$, $m_{11}^2$, and $m_{22}^2$.
We use Eq. \ref{eqn:constraint1} to \ref{eqn:constraint2} to solve for these parameters which ensures that our choice of 
parameters is one of the extrema of the potential. We then find all possible values for $(v,v_1,v_2,v_3)$ which
satisfy Eq. \ref{eqn:constraint1} to \ref{eqn:constraint2} and check to see which values lead to the smallest value of the 
potential. If the values of the vevs with which we started do not lead to the deepest minima, we must reject that point.

\section{Results and Discussion for the Hidden Gauged SU(3) Model}
\label{sec:Discussion}

The Hidden Gauged SU(3) model has too many parameters to make a full scan of the
parameter space practical.
 Therefore, we will explore the implications of three representative 
benchmark points. 
The parameter values  for  benchmark point A
were chosen to explore the region where one of the DM
 species could annihilate to SM particles through a Higgs resonance. Benchmark point B 
was chosen to explore a region where no such resonant effects would occur. Finally, benchmark point 
C was chosen to explore the region where $v_1$ and $v_2$ have similar values and where there
are many more stable DM particles.

We will vary the mass of one of the DM species and find the gauge coupling $\tilde{g}$ which 
results in the correct relic density.  We  vary the mass of $\chi$, $\mathcal{H}$, and $A^{1'}$ for 
benchmark point A, B, and C, respectively. 
Some of the hidden sector particles are stable because symmetry respecting decays
 are kinematically forbidden.
Thus, benchmark points A and B have 
4 stable DM species, while benchmark point 
C has 7 DM species. Our purpose is to  
study some of the features that arise from having many stable DM species. The benchmark points
we consider are defined in Table~\ref{table:benchmark}. For all the benchmark points we
take $\theta_2=0$ and combine $\theta_1$ and $\theta_3$ into a single angle $\theta$,  which
we set to $\sin \theta =0.1$ which is consistent with experimental
constraints coming from the SM Higgs couplings \cite{Falkowski:2015iwa,Arcadi:2016kmk}. We
also ignore any unstable dark sector species by assuming that they would have decayed out long
before freezeout occurred.

\subsection{DM Relic Density from the Hidden Gauged SU(3)  Model}
\label{sec:relic_density}

\begin{figure*}[h]
\begin{center}
$
\begin{array}{cc}
\includegraphics[scale=0.20,keepaspectratio=true]{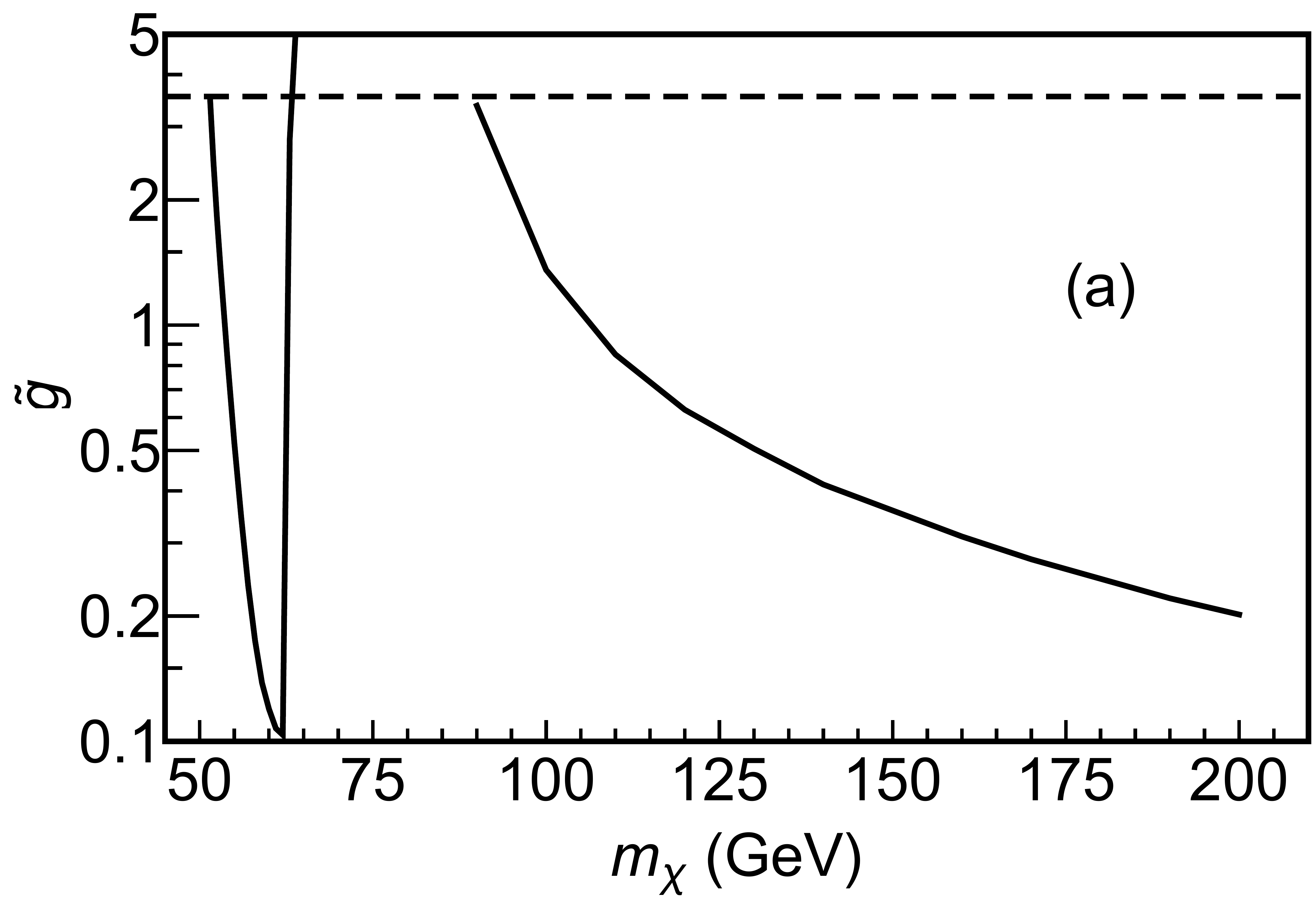} &\hspace*{0.2cm} 
\includegraphics[scale=0.20,keepaspectratio=true]{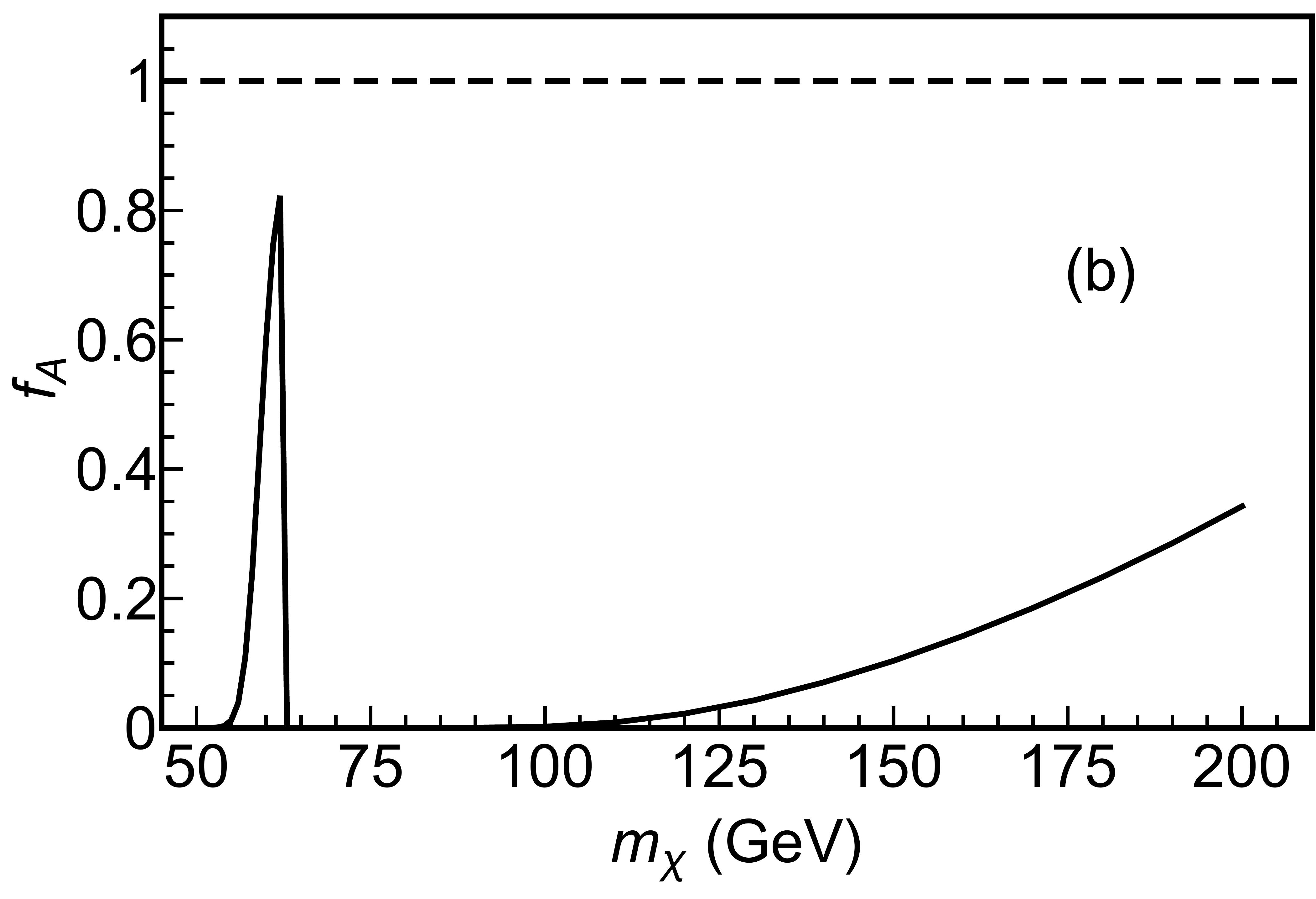}
\end{array}
$ 
\end{center}
\begin{center}
$
\begin{array}{cc}
\includegraphics[scale=0.20,keepaspectratio=true]{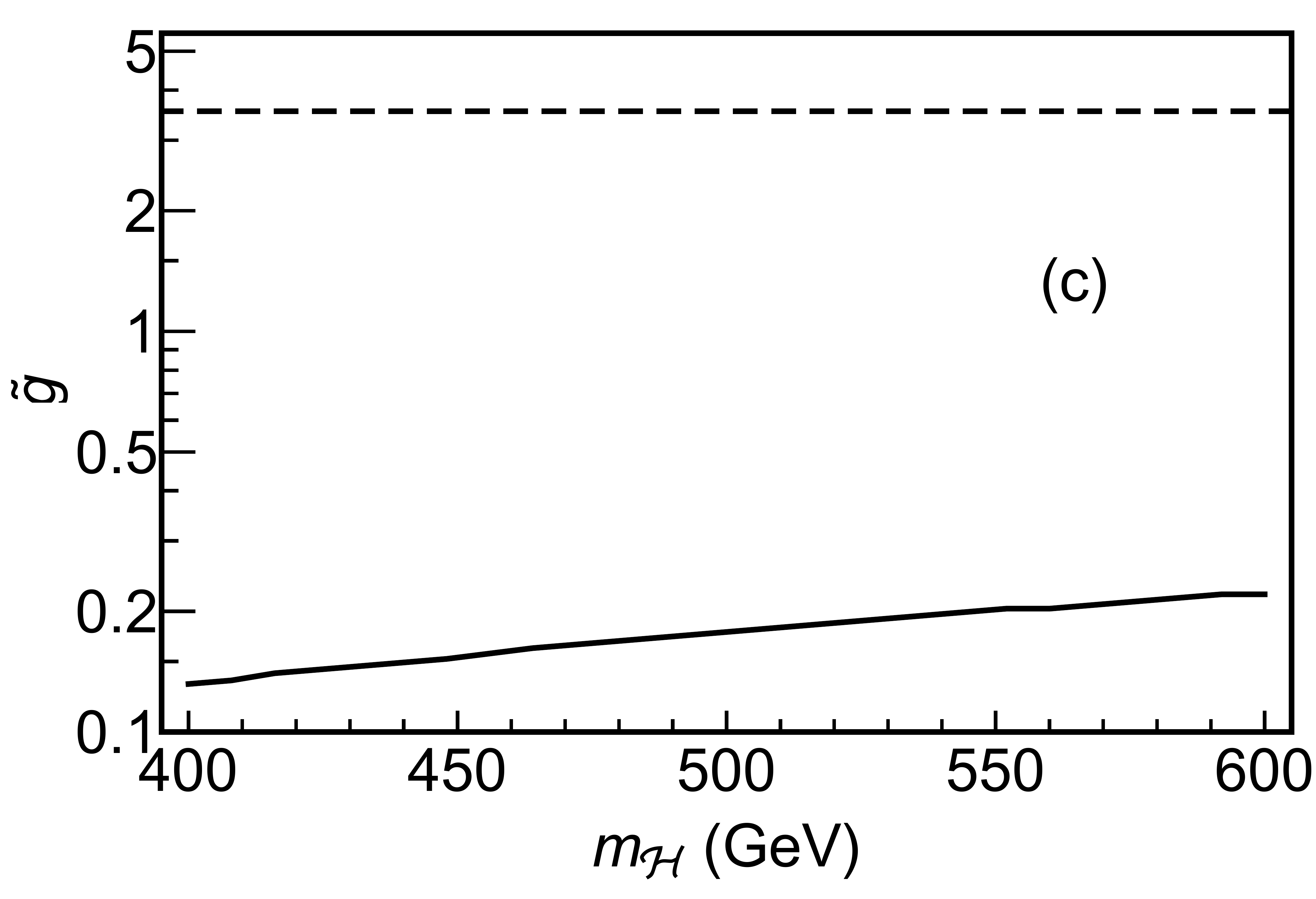} &\hspace*{0.2cm} 
\includegraphics[scale=0.20,keepaspectratio=true]{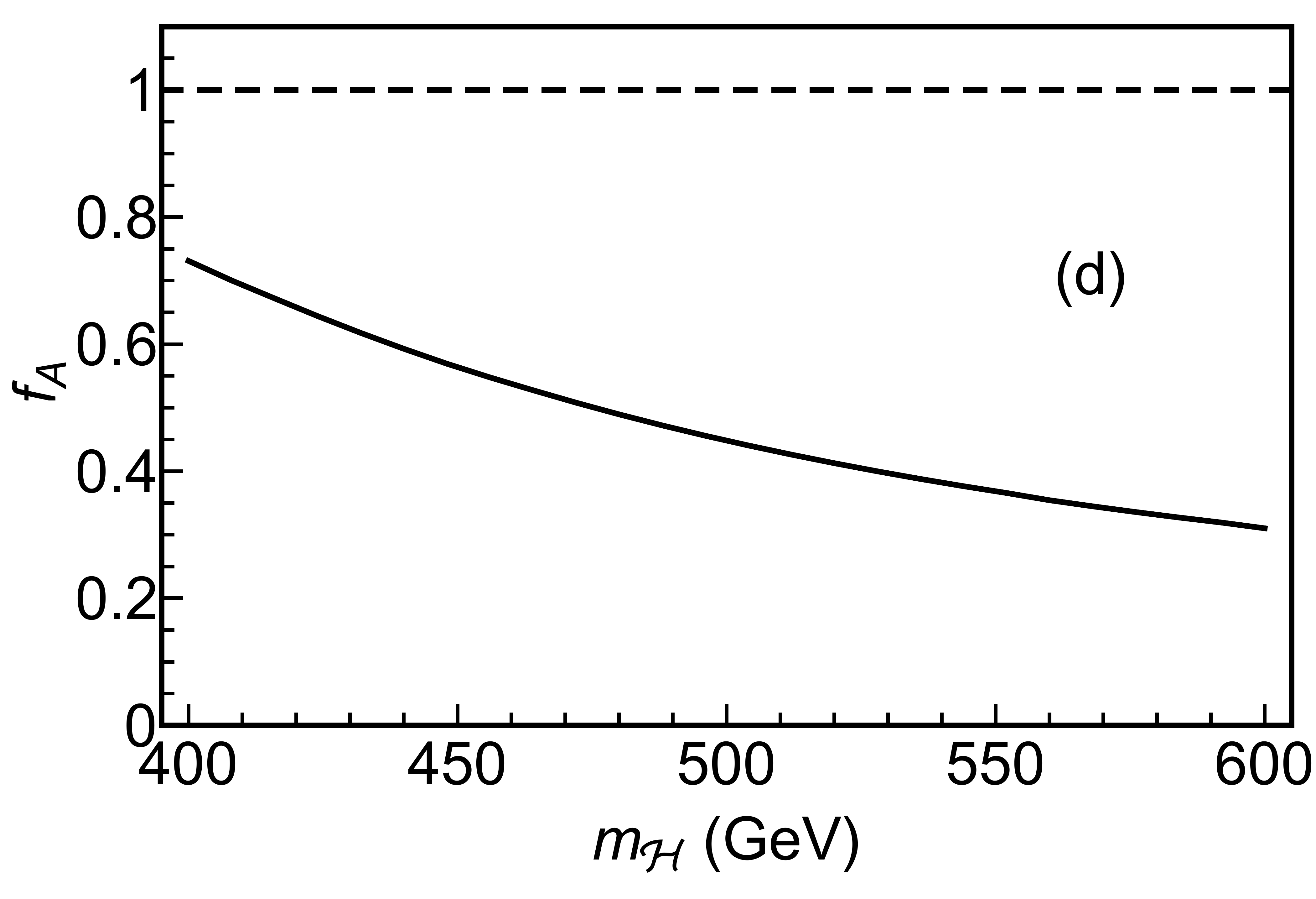}
\end{array}
$ 
\end{center}
\begin{center}
$
\begin{array}{cc}
\includegraphics[scale=0.20,keepaspectratio=true]{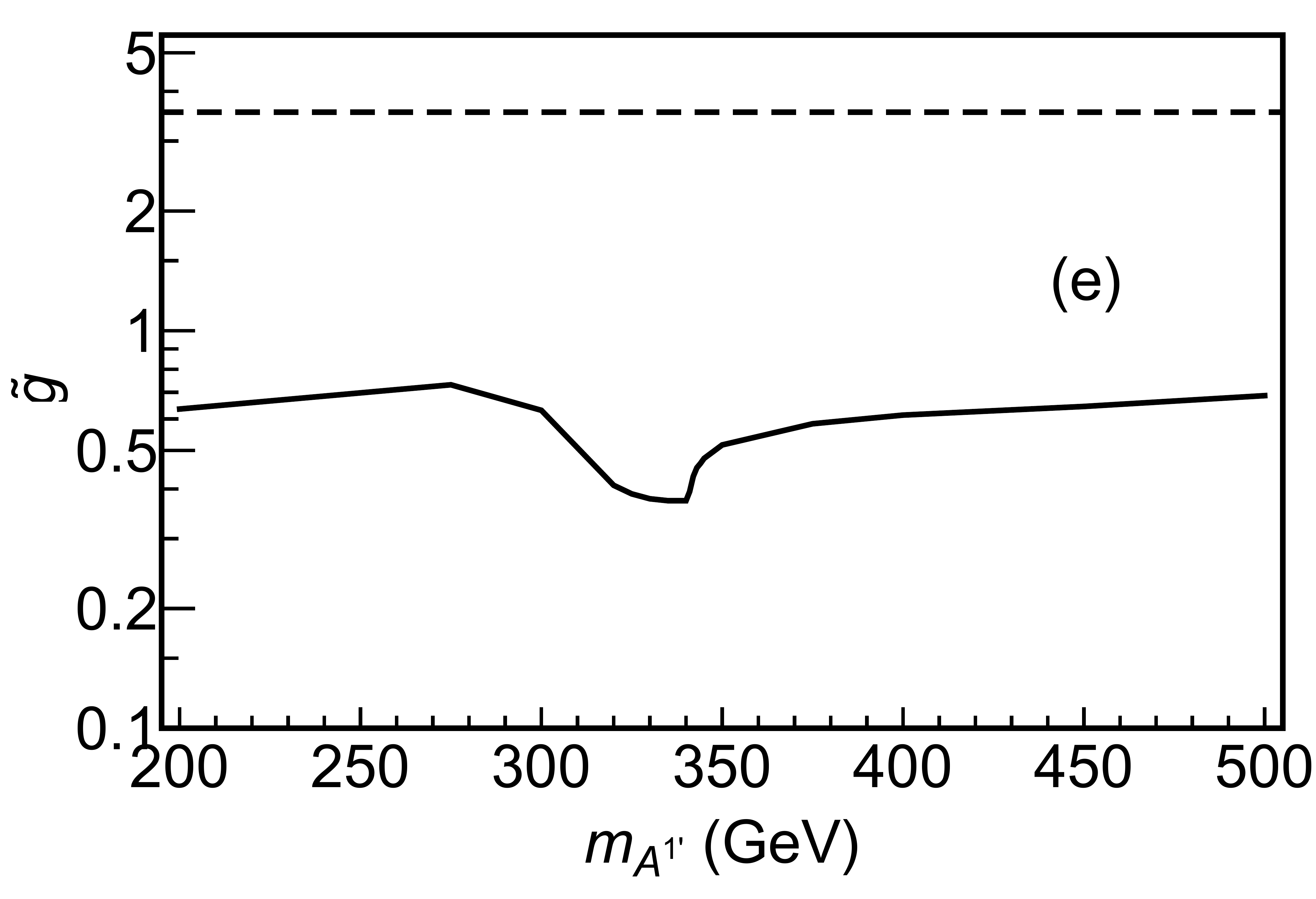} &\hspace*{0.2cm} 
\includegraphics[scale=0.20,keepaspectratio=true]{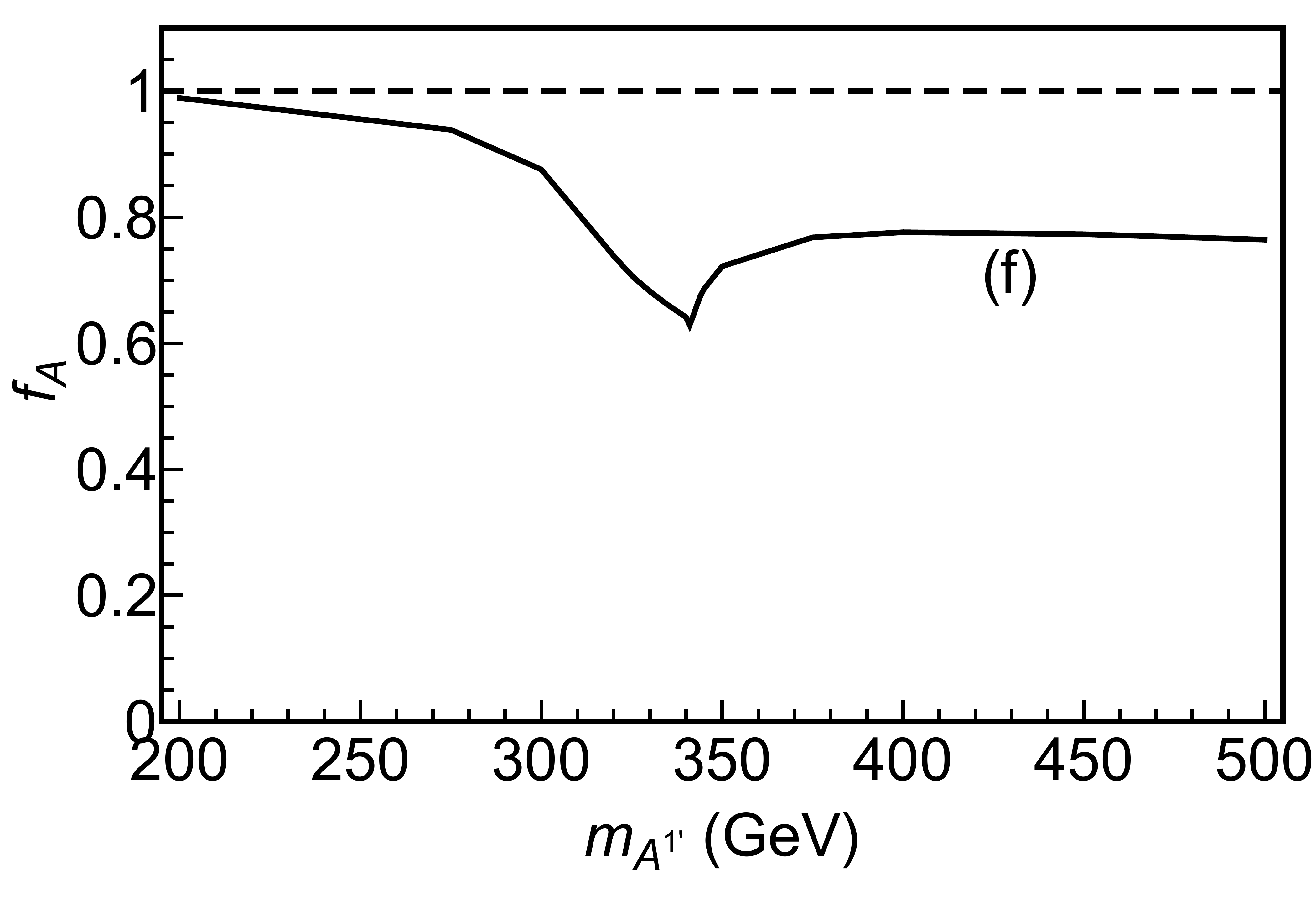}
\end{array}
$ 
\end{center}

\vskip -0.1in
   \caption{ Constraints on the allowed SU(3) gauge coupling values and the resulting
   ratio of the vector relic density to total relic density as a function of the DM mass. 
   The left column shows values of the gauge coupling required so that the relic density 
   is consistent with  $\Omega_{CDM}h^2=0.1186 \pm 0.002$, while the right column shows
   the fraction of DM that consists of vector particles.  The dashed lines in the left column show the unitarity bound 
   for $\tilde{g}$ which is $\sqrt{4\pi}$ and the dashed lines in the right column show the maximum value that $f_A$ can take 
   which is 1. Plots (a) and (b) show the results for scenario A, plots (c) and (d) show the results for scenario B, and plots (e)
   and (f) show the results for scenario C.
}
\label{fig:gfCompare}
\end{figure*}

When varying the mass of one of the DM species for a benchmark point, we used a binary search to find
a value of the gauge coupling which would result in the relic density lying in the measured range of
$\Omega_{CDM}h^2=0.1186 \pm 0.002$ \cite{Steigman} . 
Figure~\ref{fig:gfCompare} shows the resulting values for
the coupling as a function of the DM mass for each benchmark point.
If the required value of the coupling fell outside the 
unitarity bound of $\sqrt{4\pi}$, we did not consider it. 
This was the case for benchmark point A
for a range of $m_\chi$ masses above the
threshold for  annihilation via the Higgs boson  resonance
 as can be seen by the lack of points between half the Higgs
mass and about $90$ GeV.

Another insightful variable is the ratio of the relic density of vector particles to the total
relic density. This is defined by
\begin{align}
f_A=\frac{\sum_i \Omega_{{A^i}'}}{\Omega_{\rm tot}}.
\end{align}
This variable shows how the composition of the DM changes as we vary 
parameters such as the gauge coupling. It is very useful when looking at direct and indirect detection since
the scalar and vector particles interact very differently with detectors. Figure \ref{fig:gfCompare}
shows how  $f_A$ varies for each benchmark points.


\subsection{DM Direct Detection}
\label{sec:DM_direct_constraint}
We now examine the constraints obtained from direct detection measurements using the latest results from the XENON1T 
experiment
\cite{XENON1T:2017,XENON1T:2018}. Because there are multiple DM species that can each interact differently
with the detector, we cannot simply calculate a cross section and compare it with the reported experimental 
bound. Instead, we must find the predicted 
number of events that the experiment would have seen,
 and compare that to what the experiment observed. The theoretical rate of events is given by 
\begin{align}
\frac{dR}{dE_R}=\sum_{i} f_i \left(\frac{dR}{dE_R}\right)_i,
\end{align}
where $f_i=\frac{\Omega_i}{\Omega_{\rm tot}}$ and \cite{KeithBrooks:2012}
\begin{align}
\left(\frac{dR}{dE_R}\right)_i = \frac{\sigma_{iN}^0 \rho_i^{\rm loc}}{2\mu_{iN}^2 m_i}F_i^2(E_R)I_i(E_R).
\end{align}
Here, $E_R$ is the nuclear recoil energy in the detector, $\sigma_{iN}^0$ is the DM nucleus cross section at zero momentum 
transfer, $\rho_i^{\rm loc}$ is the local DM density for the ith species, $\mu_{iN}=m_i m_N/(m_i+m_N)$ 
is the reduced mass of the nucleus DM system, $F_i(E_R)$ is a nuclear form factor, and $I_i(E_R)$ 
is the mean of the inverse of the speed, $\langle v_i^{-1} \rangle $, 
of the ith species in the DM halo for a given $E_R$. Explicit formulas
for $I_i(E_R)$ and $F_i(E_R)$ can be found in Ref.~\cite{KeithBrooks:2012}. Once the differential rate is calculated,
we integrate it over the  energy range appropriate for the experiment and multiply the result by an efficiency factor which
could in principle be different for each species. In the case for XENON1T, the energy range for single nuclear
recoil events is 4.9 to 40.9~keV, and we assume a detection efficiency of $89\%$, which for simplicity we took to be the same
for all DM species and recoil energies.

For the parameter values to be allowed, the calculated theoretical rate should fall below the experimental limit. At a
confidence level of $1-\beta$, a background rate $\nu_b$, and a total of $n_{\rm obs}$ observed events,
we can find the upper limit on the signal rate $\nu_s^{\rm up}$ by numerically solving:
\begin{align}
\beta=\frac{e^{-(\nu_s^{\rm up}+\nu_b)}\sum_{n=0}^{n_{\rm obs}} \frac{(\nu_s^{\rm up}+\nu_b)^n}{n!}}{e^{-\nu_b}\sum_{n=0}^{n_{\rm obs}}\frac{\nu_b^n}{n!}}.
\end{align}
For XENON1T \cite{XENON1T:2017,XENON1T:2018}, 
we use $\nu_b=735$ and $n_{\rm obs}=739$. To compare the experimental rate  to
the theoretical prediction, the experimental rate needs to be given in terms of events per unit
time per unit mass which is achieved by dividing the experimental values by the exposure. For XENON1T, this is
$1.0\mbox{ t}\times\mbox{yr}$.
At the 95\% confidence level, the resulting limit on the rate is 
$1.58\times 10^{-4}\mbox{ kg}^{-1}\mbox{day}^{-1}$. \cite{XENON1T:2018}.

\begin{figure}[h]
\begin{center}
$
\includegraphics[scale=0.18,keepaspectratio=true]{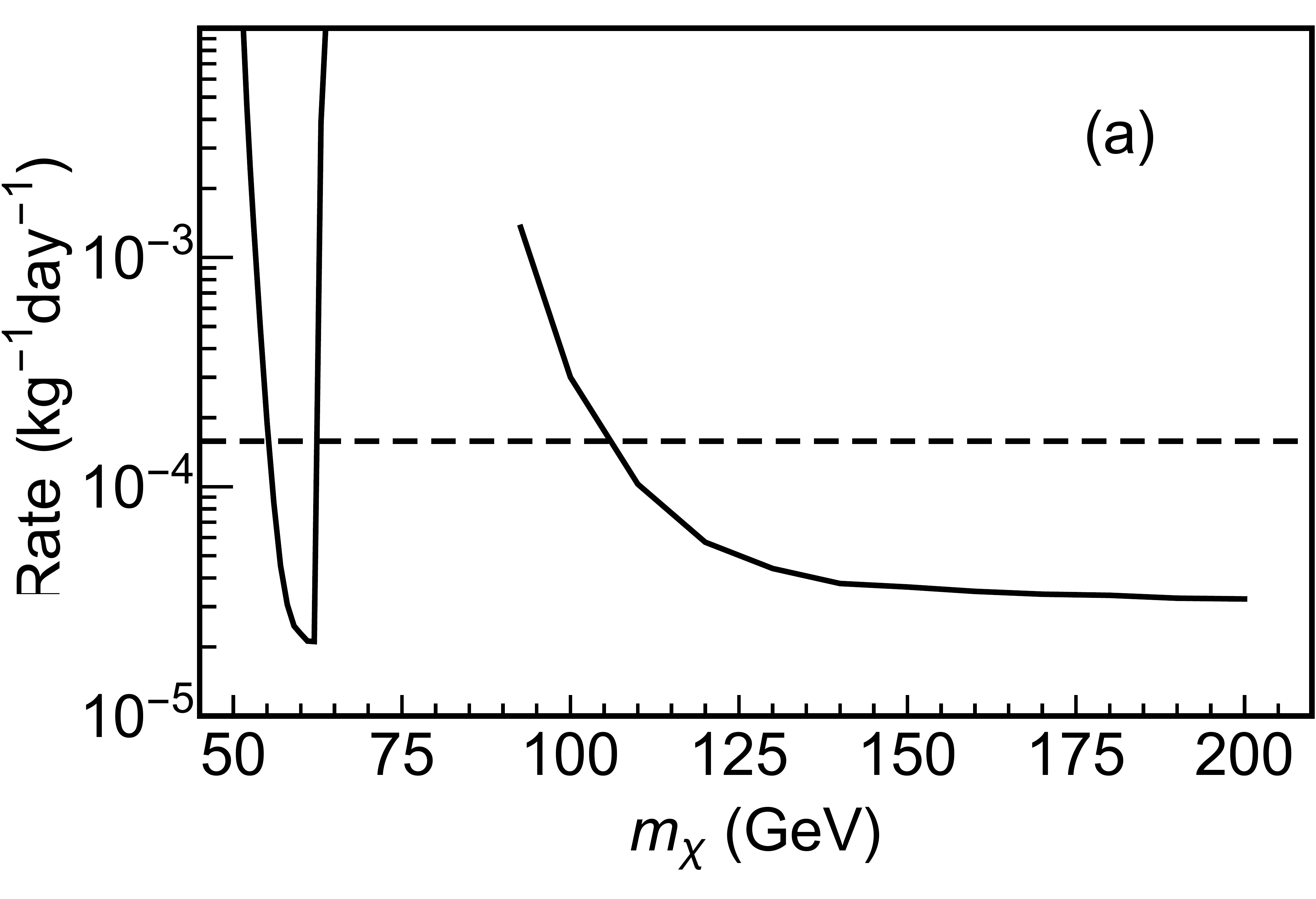} 
$ 
\end{center}
\begin{center}
$
\includegraphics[scale=0.18,keepaspectratio=true]{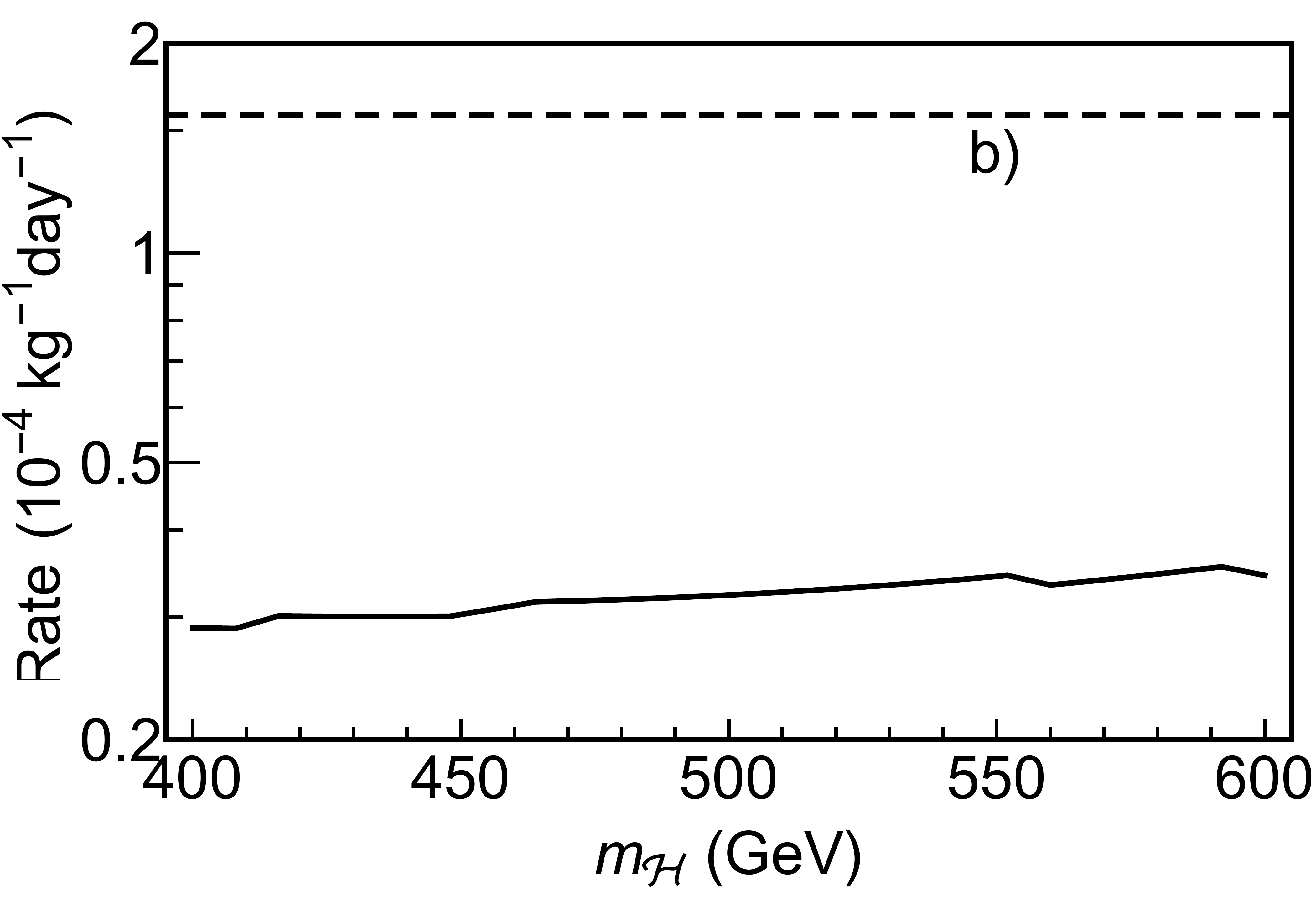}
$ 
\end{center}
\begin{center}
$
\includegraphics[scale=0.185,keepaspectratio=true]{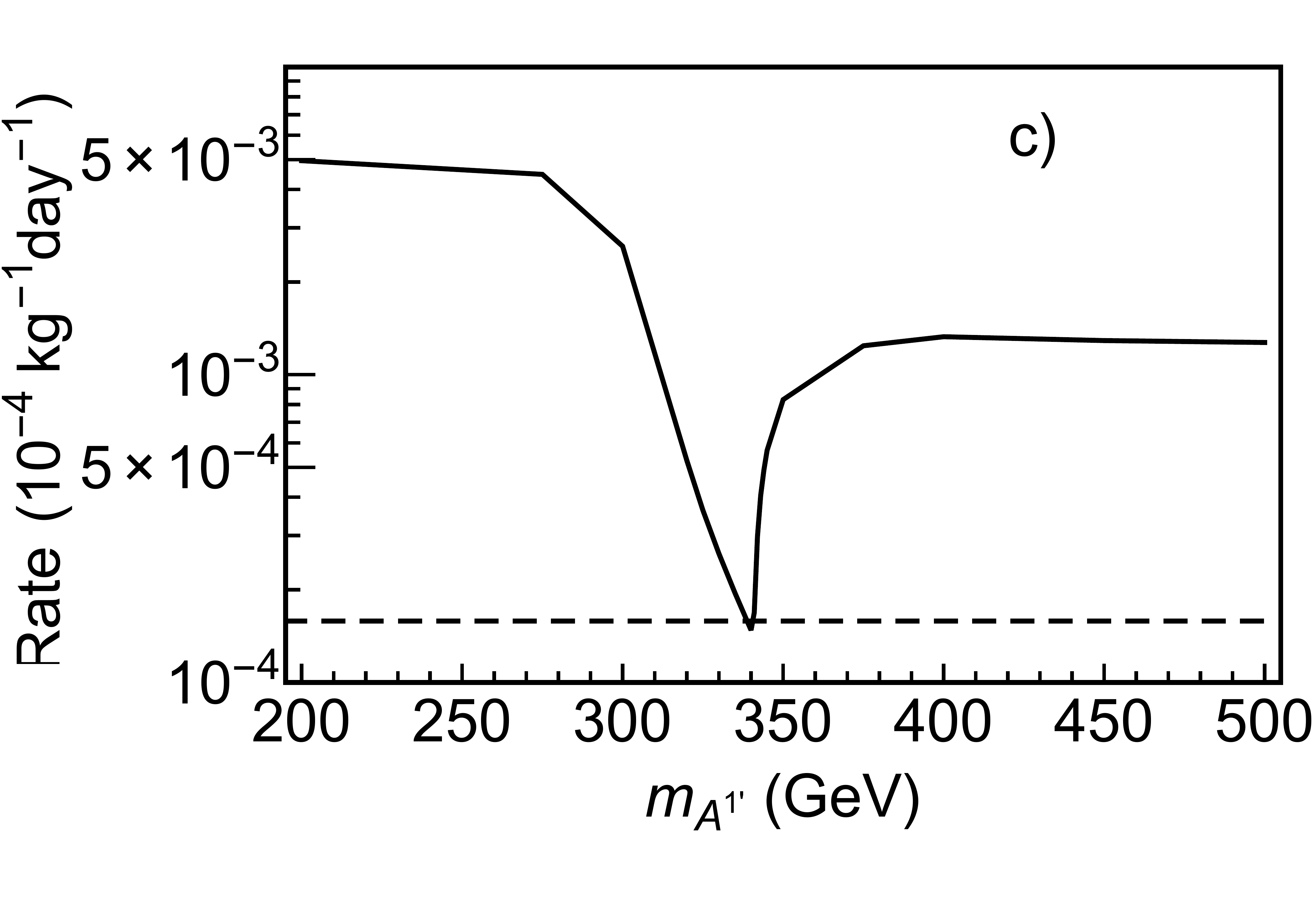}
$ 
\end{center}
%
\vskip -0.1in
   \caption{ The direct detection event rate as a function of the DM mass. 
   Plots (a), (b), and (c) show the results for benchmark point A, B and C, respectively.
   The allowed points are below the dashed line at $1.58\times 10^{-4}\mbox{ kg}^{-1}\mbox{day}^{-1}$.
   In all the cases, the regions where the coupling constant $\tilde{g}$ was large to give the allowed 
   relic density are ruled out by these direct detection constraints. These are also the regions where
   a significant portion of the DM is comprised of scalars.
}
\label{fig:DDab}
\end{figure}

For the cross section, our benchmark points have two mediating particles between the dark and visible sectors, 
namely $h_1$ and $h_3$. The cross section for DM species $i$ and a nucleon is thus given by:
\begin{align}
\sigma_{in}^0=\frac{f_N^2}{4\pi}\left(\frac{g_{iih_1}c_\theta}{m_{h_1}^2}+\frac{g_{iih_3}s_\theta}{m_{h_3}^2}\right)^2\frac{m_n^2}{(m_n+m_i)^2v^2},
\label{eqn:dd}
\end{align}
where $c_\theta$ and $s_\theta$ are $\cos\theta$ and $\sin\theta$, respectively, $f_N$ is the SM 
Higgs effective coupling to nucleons which is approximately $0.30 \pm 0.03$ \cite{Cline:2013}, $m_n$ is the nucleon 
mass which we take to be $0.93895$ GeV, $m_i$ is the mass of the DM species in question, $v$ is the
SM vev, and $g_{iih_1}$ and $g_{iih_3}$ are the couplings between the DM and $h_1$ and $h_3$,
respectively. In the case of scalar DM, we take the Feynman rule to be of the form $-ig$, and in the
case of  vector DM, the Feynman rule is of the form $-igg_{\mu\nu}$. To transform Eq.~\ref{eqn:dd} to 
a cross section with the entire nucleus, we use \cite{KeithBrooks:2012}:
\begin{align}
\sigma_{iN}^0=\sigma_{in}^0\frac{A^2\mu_{iN}^2}{\mu_{in}^2},
\end{align}
where $A$ is the atomic number of the nucleus, $\mu_{iN}$ is the reduced mass between the nucleus 
and the DM 
particle, and $\mu_{in}$ is the reduced mass between the nucleon and the DM particle. 
The scalar DM particles ($\chi$ and ${\cal H}$) interact weakly with the nuclear target.  This
is because the scalar-$h_1$ coupling is weak due to our choice of $\lambda_{H11}=0$ and $\lambda_{H22}$ to be small
 and $h_3$ interacts
with nucleons via its SM Higgs component which is small (recall $\sin\theta =0.1$).  As a consequence, when DM is
mainly scalar it is weakly constrained by direct detection.
The results for each benchmark point are shown in Fig~\ref{fig:DDab}.  Again,  the gap between half the Higgs 
mass and roughly 
90 GeV for benchmark point A
is because the value of the coupling that gives the correct relic abundance  is not allowed 
because it violates the 
unitarity constraint. In Fig~\ref{fig:DDab}(a), we see that there is an allowed region at high mass and 
in the Higgs resonance region.
In Fig~\ref{fig:DDab}(b), the gauge coupling is quite small and has a large fraction of scalar DM so 
direct detection does not impose any significant constraints on benchmark point B. In Fig~\ref{fig:DDab}(c), 
we see that direct detection excludes most of the mass region 
 due to the large vector 
fraction and large gauge coupling. There is however the tip of a resonance which allows for efficient 
annihilation of vectors around $340$ GeV which remains allowed. This corresponds to when the mass 
of $A^{3'}$ is just below $300$ GeV and where there is significant annihilation of $A^{3'}$ 
through the $600$ GeV $h_2$.

We see that $f_A$ is non-trivial in each case, that is to say, it is not close to being exactly 1 or exactly 0. This
non-trivial behavior tells us that it is important to consider all the relevant particles when doing the freeze-out calculation, 
not just the lightest or most weakly coupled. In the cases where a resonance effect is present, it is particularly important
to be precise, as these regions require quite different couplings constants to obtain the observed relic density.

\subsection{DM Indirect Detection}
\label{sec:DM_indirect_constraint}

Dwarf spheroidal satellite galaxies (dSphs) are typically DM dominated so are 
a good place to study DM \cite{Slatyer:2017sev,Conrad:2015bsa}. 
The Fermi collaboration \cite{FermiLAT} has acquired  6 years of data from
 observing 15 dSphs and have released bounds for WIMP DM annihilation based on 
the observed gamma ray flux. They considered the following representative final states for the 
DM annihilation: $e^+e^-$, $\mu^+\mu^-$, $\tau^+\tau^-$, $u\bar{u}$, $b\bar{b}$, and $W^+W^-$ \cite{FermiLAT}.

\begin{figure}[H]
\begin{center}
$
\includegraphics[scale=0.20,keepaspectratio=true]{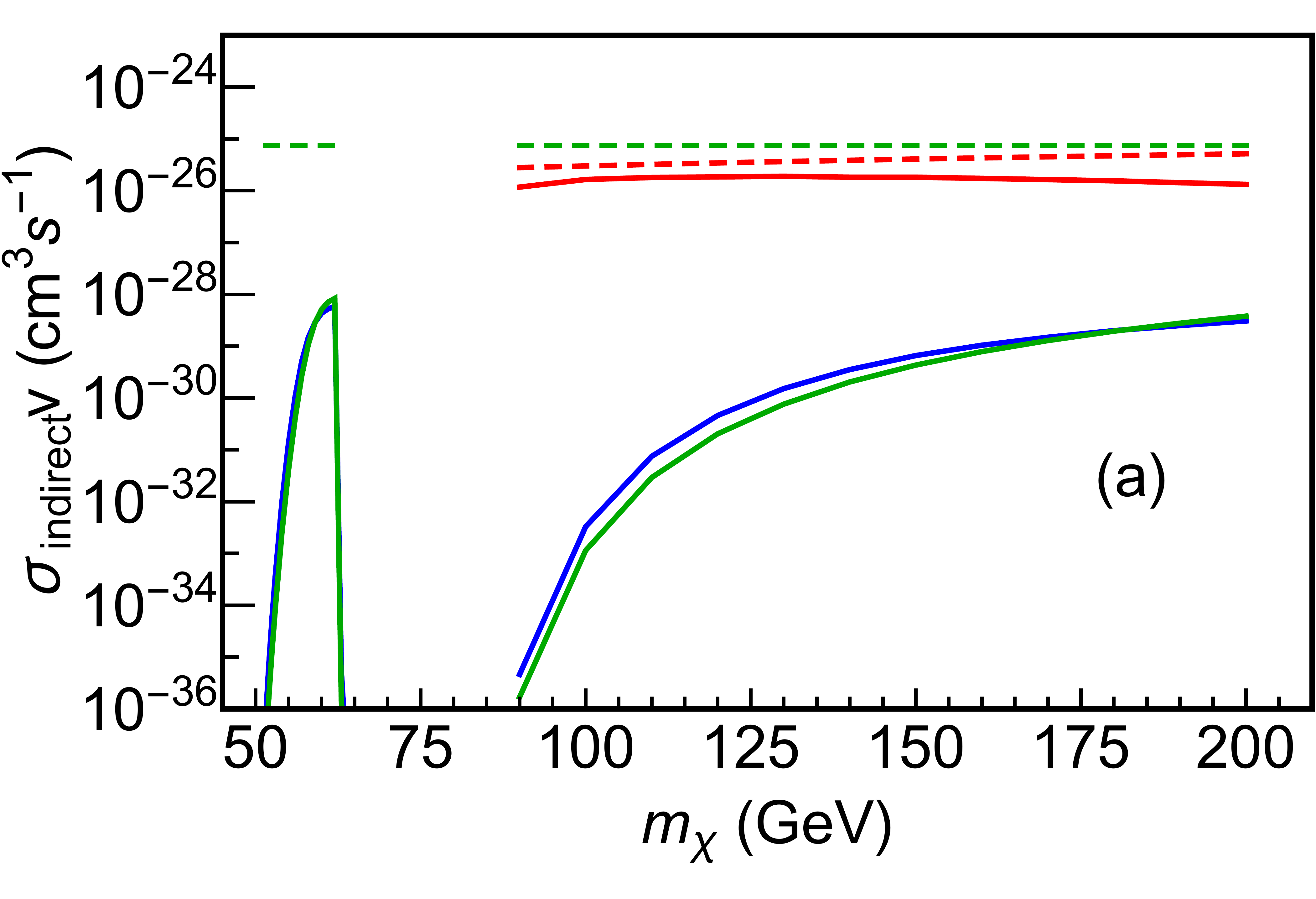} 
$
\end{center}
\begin{center}
$
\includegraphics[scale=0.20,keepaspectratio=true]{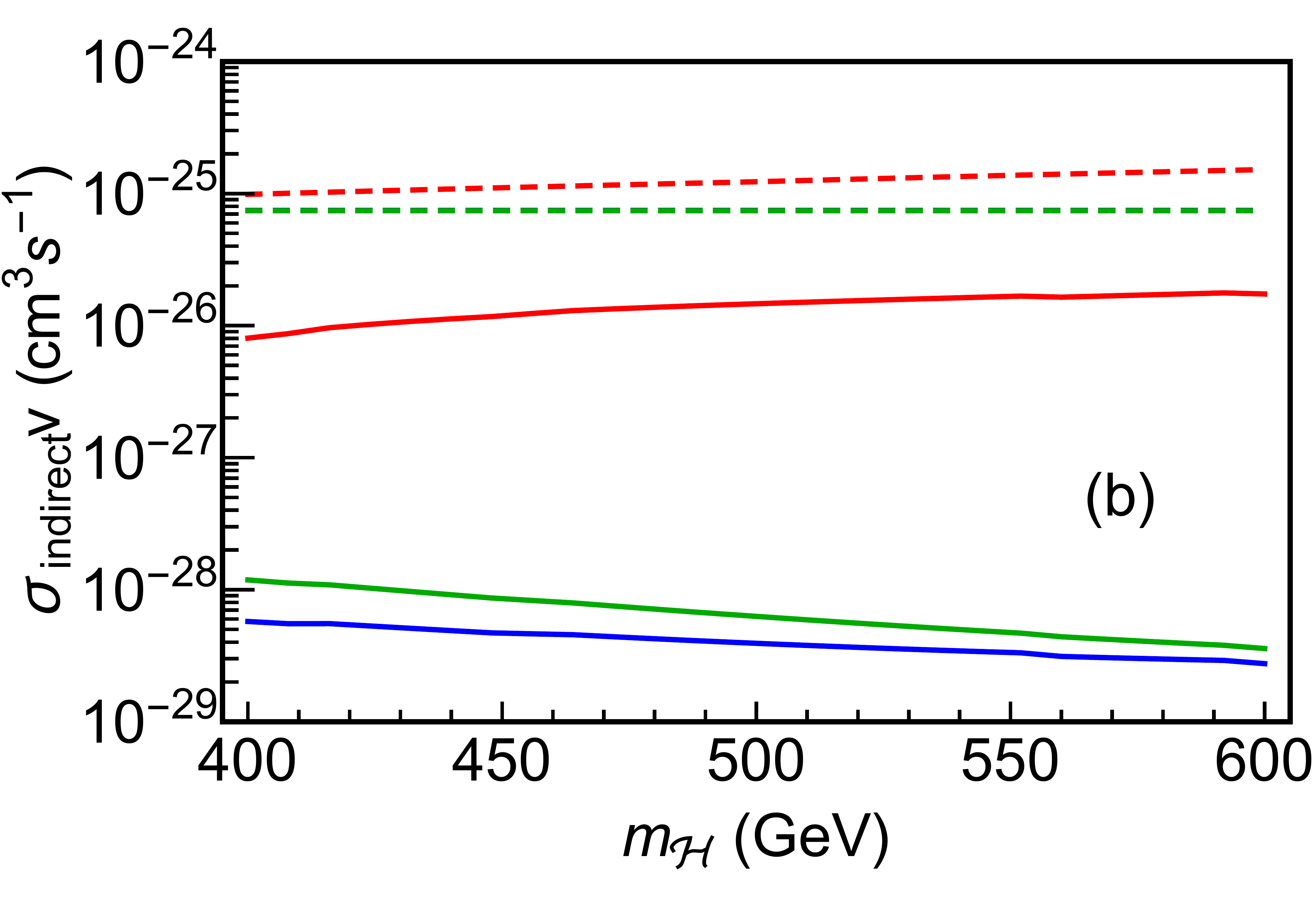}
$ 
\end{center}
\begin{center}
$
\includegraphics[scale=0.204,keepaspectratio=true]{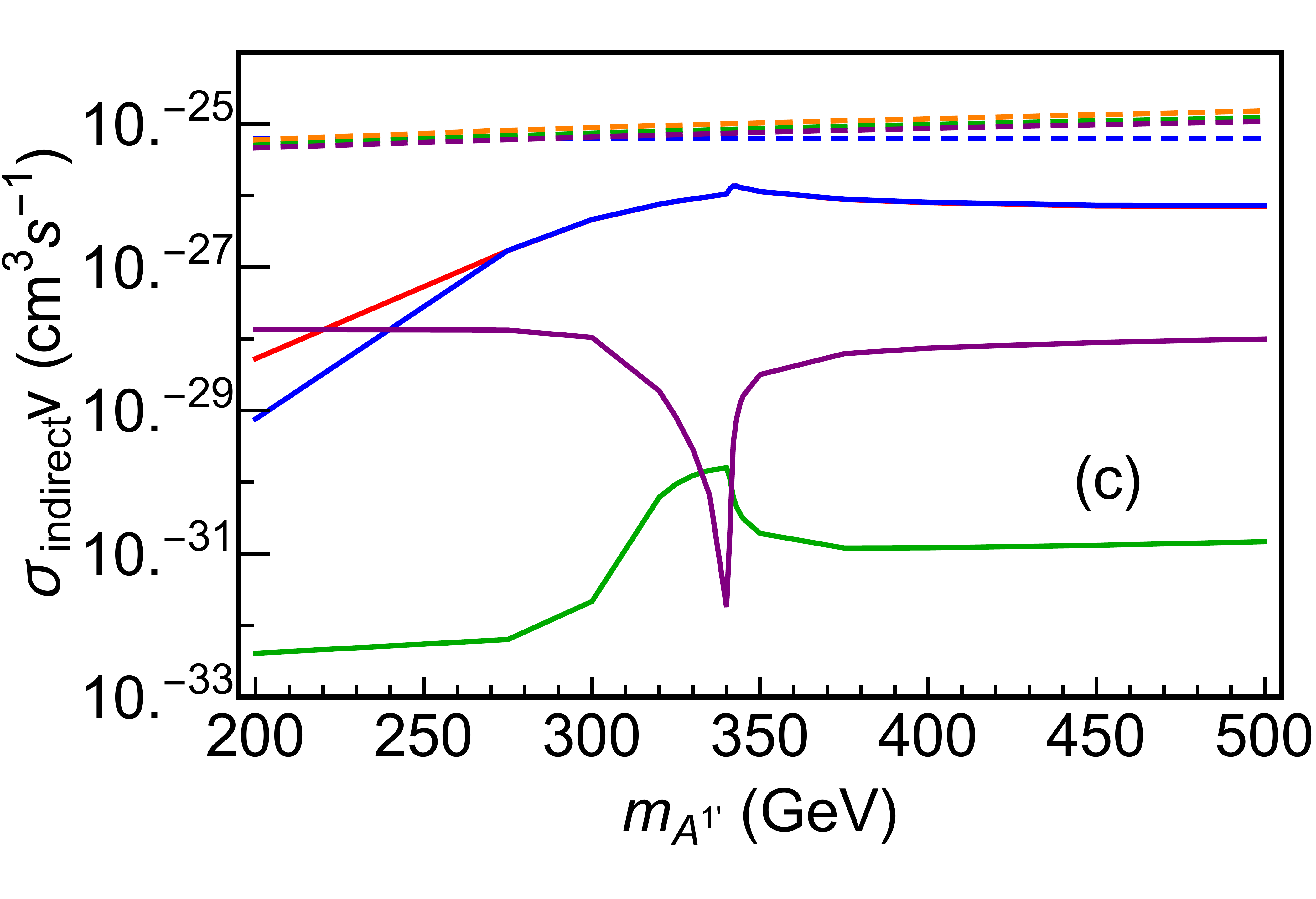}
$ 
\end{center}
\vskip -0.1in
   \caption{ 
   Indirect detection constraints for the $W^+W^-$ final state. Figures (a), (b), and (c) show the results
   for benchmark point A, B, and C, respectively.
   The  line colors represent:
   red 
   for the $\chi\chi\rightarrow W^+W^-$, blue  
   for $\mathcal{H}\mathcal{H}\rightarrow W^+W^-$, green for $A^{1'}A^{1'}\rightarrow W^+W^-$ or $A^{2'}A^{2'}\rightarrow W^+W^-$ which
   have identical results, purple 
   for $A^{3'}A^{3'}\rightarrow W^+W^-$, and orange for $A^{4'}A^{4'}\rightarrow W^+W^-$ or $A^{5'}A^{5'}\rightarrow W^+W^-$ which
   have identical results. The dashed lines are the constraints from Fermi \cite{FermiLAT}
   while the solid lines are the predictions 
   of the model. Note that the purple and green dashed line are almost superimposed because they have 
   very similar masses leading to similar constraints. We also note that the solid orange line does not appear
   because it is too small.
}
\label{fig:indirectRes}
\end{figure}

We translate these bounds into constraints on our model by considering the cross sections
of pairs of identical DM particles into the above final states scaled by their fraction
squared and their branching ratio.  Explicitly, we will compare:
\begin{align}
\sigma_{\rm scaled}=\left(\frac{\Omega_i}{\Omega_{DM}}\right)^2 
{ { (\sigma_{ii\rightarrow SM+SM} )^2 } \over{ \sigma_{ii}^{tot} }},
\end{align}
where $\sigma_{ii\rightarrow SM+SM}$ is the cross section from two identical DM particles
to one of the $SM$ final states considered by the Fermi collaboration and 
$\sigma_{ii}^{tot} $ is the sum of the cross sections to all possible final states.
Of all the final states considered by Fermi, only the $W^+W^-$ final state has the potential 
to provide useful constraints as the scaled cross sections to the other final states are orders of magnitude 
below the Fermi constraints. This is because the cross sections to these fermions are 
proportional to the mass squared of the fermion which are small compared to the $W$ mass squared. 
Fig.~\ref{fig:indirectRes} shows the
results for each of the benchmark points. In all cases, the scalar DM ends up having a significantly larger scaled
cross section to $W^+W^-$ than any vector DM because the vector DM annihilates preferentially 
to other DM particles. 
We see that although the constraint is close,  indirect detection does not impose any constraints. 
For benchmark point A the experimental limits are quite close to the theoretical predictions while
for benchmark points B and C the experimental limits are about an order of magnitude above the theoretical
predictions.  Thus, in all three cases there is the potential to rule out large regions of
parameter space with moderate improvements to the experimental bounds. 
It should also be noted that a more sophisticated analysis that 
considers the resulting spectrum from multiple
DM species co-annihilating to the same final state could result in more stringent constraints.

These results lead to the interesting observation that if the DM is dominated by vector species, then
it will be more easily detected in direct detection experiments while if the DM is dominated by
scalar species, then it would more likely to be detected by indirect detection experiments. This could lead to
an observation in one type of experiment, but no signal in the other and emphasizes the importance
of the complementarity of multiple types of searches.  Arcadi {\it et al.} \cite{Arcadi:2016kmk} 
and many others have also pointed this out, 
e.g.  \cite{Gelmini:2015zpa,Bertone:2004pz,Feng:2010gw,Roszkowski:2017nbc,Arcadi:2017kky,KeithBrooks:2012,Dienes:2014via,Dienes:2017ylr}.

To end this section, we note that the constraints used are conservative. 
We only considered constraints from the 
annihilation of one species at a time. 
This bound could be improved with a more detailed study of the gamma-ray spectrum \cite{Boddy:2016fds,Boddy:2016hbp}
or cosmic-ray spectra \cite{Dienes:2013xff}
resulting from multiple species annihilating or co-annihilating and comparing this to 
the observed spectrum from dSphs \cite{Elor:2015bho}. 

\section{Summary and Conclusions}
\label{sec:conclusions}

This paper reported on a study of multi-species DM, in particular, the Hidden Gauged SU(3) model.  We  
calculated the DM relic density to constrain the parameters of the model.  
Using the resulting allowed values of the parameters 
we studied the prospects for observing multi-species DM in direct detection and 
indirect detection measurements.  Before examining the Hidden Gauged SU(3) model, we 
studied the relic abundance of simplified scenarios to gain some insight into how adding
different types of interactions between the DM species would affect the relic abundance.
We then examined a  specific model,  the Hidden Gauged SU(3) model of Arcadi {\it et al.} \cite{Arcadi:2016kmk}.
With many DM particles, it was not possible to implement a detailed scan of the parameter space.
Instead, we studied three representative benchmark points.  In all cases, we fixed the
gauge coupling to give the observed relic density.  When the relic density was dominated
by vector particles, we found that it could more easily be constrained by direct detection experiments
while if the relic abundance was dominated by scalars, indirect detection has the potential to be more constraining in the future, 
even with conservative bounds. This
reinforces the need to search for DM with a broad experimental program.  An important result was that for the different
benchmark points, the required value for the gauge coupling could vary quite significantly based on the choice of parameter values.
This can lead to significantly different relic abundances among particle species which in turn, changes the detection prospects of 
a particular parameter point. We also saw that for each benchmark point, the ratios of vectors to scalars was never dominated by either 
scalars or vectors and that the DM consisted of some non-trivial mixture of both. This shows the importance of not only including the 
lightest stable particle in the freeze-out calculation but to include all relevant particles.

\begin{acknowledgments}
The authors thank Keith Dienes and Brooks Thomas for introducing them to the
subject of DM.  They especially appreciate Keith and Brooks making available
an advance copy of their upcoming paper \cite{DHT2018}, 
for many helpful and
illuminating conversations concerning multi-species DM which inspired this work, and
for constructive comments.    
The authors also thank Heather Logan for helpful conversations. 
SG thanks the COEPP nodes of the  University of 
Melbourne, Monash University and the University of Adelaide for their hospitality where
some of this work took place.
This work was supported by the Natural Sciences and Engineering Research Council of Canada.  
\end{acknowledgments}

\appendix
\section{Details of the $\lambda_4=-\lambda_5$ Case}
\label{lambda}

For completeness we give the details of the   $\lambda_4=-\lambda_5$ case below. 
In this case, the unitary gauge is given by:
\begin{widetext}
\begin{align}
H=&\frac{1}{\sqrt{2}}\begin{pmatrix}
0\\ h+v
\end{pmatrix},\notag\\ 
\Phi_1=&\frac{1}{\sqrt{2}}\begin{pmatrix}
0\\ 
\frac{v_1v_2\varphi_3 }{\sqrt{(v_1^2+v_2^2)(v_1^2+v_2^2+v_3^2)}} -\frac{v_3\varphi_2 }{\sqrt{v_1^2+v_3^2}}-i\frac{v_2\varphi_1}{\sqrt{v_1^2+v_2^2+v_3^2}}\\
v_1+\varphi_4-i\frac{v_3\varphi_1}{\sqrt{v_1^2+v_2^2+v_3^2}}
\end{pmatrix},\
\Phi_2=\frac{1}{\sqrt{2}}\begin{pmatrix}
0\\ 
v_2+\frac{v_1\varphi_2 }{\sqrt{v_1^2+v_3^2}}-\frac{v_2v_3\varphi_3 }{\sqrt{(v_1^2+v_2^2)(v_1^2+v_2^2+v_3^2)}} \\
v_3+\sqrt{\frac{v_1^2+v_3^2}{v_1^2+v_2^2+v_3^2}}\varphi_3+i\frac{v_1\varphi_1}{\sqrt{v_1^2+v_2^2+v_3^2}}
\end{pmatrix}.\
\end{align}
We write down the scalar mass matrix with $\lambda_4=-\lambda_5$ in the form $\mathcal{L}\supset \frac{1}{2}\Phi^T m^2_{\rm scalar}\Phi$ where $\Phi^T=(h, \varphi_4, \varphi_2, \varphi_3, \varphi_1)$ are the scalars in the model, and where 
\begin{align}
&m^2_{\rm scalar}=\notag\\
&\begin{pmatrix}
\lambda_H v^2 & \lambda_{H11}v v_1&  \frac{\lambda_{H22}v v_1 v_2}{\sqrt{v_1^2+v_3^2}} &   \frac{\lambda_{H22}v v_3(v_1^2 - v_2^2 + v_3^2)}{\sqrt{(v_1^2 + v_3^2)(v_1^2 + v_2^2 + v_3^2)}}  & 0\\
\lambda_{H11}v v_1& \lambda_1 v_1^2 & \frac{\lambda_{3}v_1^2 v_2}{\sqrt{v_1^2+v_3^2}} & \frac{\lambda_{3}v_1 v_3(v_1^2 - v_2^2 + v_3^2)}{\sqrt{(v_1^2 + v_3^2)(v_1^2 + v_2^2 + v_3^2)}} & 0\\
 \frac{\lambda_{H22}v v_1 v_2}{\sqrt{v_1^2+v_3^2}} & \frac{\lambda_{3}v_1^2 v_2}{\sqrt{v_1^2+v_3^2}} &  \frac{\lambda_{2}v_1^2 v_2^2}{v_1^2+v_3^2} & \frac{\lambda_{2}v_1 v_2 v_3(v_1^2 - v_2^2 + v_3^2)}{(v_1^2 + v_3^2)\sqrt{(v_1^2 + v_2^2 + v_3^2)}} & 0\\
\frac{\lambda_{H22}v v_3(v_1^2 - v_2^2 + v_3^2)}{\sqrt{(v_1^2 + v_3^2)(v_1^2 + v_2^2 + v_3^2)}}&\frac{\lambda_{3}v_1 v_3(v_1^2 - v_2^2 + v_3^2)}{\sqrt{(v_1^2 + v_3^2)(v_1^2 + v_2^2 + v_3^2)}}& \frac{\lambda_{2}v_1 v_2 v_3(v_1^2 - v_2^2 + v_3^2)}{(v_1^2 + v_3^2)\sqrt{(v_1^2 + v_2^2 + v_3^2)}} & \frac{ \lambda_{2}v_3^2(v_1^2 - v_2^2 + v_3^2)^2}{(v_1^2 + v_3^2)(v_1^2 + v_2^2 + v_3^2)}&0\\
0&0&0&0& \lambda_4(v_1^2+v_2^2+v_3^2)
\end{pmatrix}.
\end{align}
\end{widetext}

This case also results in more mixing between the gauge bosons. They are described by the following mass terms:

\pagebreak
\begin{align}
\mathcal{L}&\supset
\frac{\tilde{g}^2}{8}\begin{pmatrix}
A^1_\mu \\ A^4_\mu
\end{pmatrix}^T
\begin{pmatrix}
v_2^2 & v_2v_3\\
v_2v_3 & v_1^2+v_3^2
\end{pmatrix}
\begin{pmatrix}
A^{1\mu} \\ A^{4\mu}
\end{pmatrix}\notag\\&+
\frac{\tilde{g}^2}{8}\begin{pmatrix}
A^2_\mu \\ A^5_\mu
\end{pmatrix}^T
\begin{pmatrix}
v_2^2 & v_2v_3\\
v_2v_3 & v_1^2+v_3^2
\end{pmatrix}
\begin{pmatrix}
A^{2\mu} \\ A^{5\mu}
\end{pmatrix}\notag\\&+
\frac{\tilde{g}^2}{8}\begin{pmatrix}
A^3_\mu \\A^6_\mu \\ A^8_\mu
\end{pmatrix}^T
\begin{pmatrix}
v_2^2 & -v_2v_3 & \frac{-v_2^2}{\sqrt{3}}\\
-v_2v_3 & v_1^2+v_2^2+v_3^2&\frac{-v_2v_3}{\sqrt{3}}\\
\frac{-v_2^2}{\sqrt{3}} &\frac{-v_2v_3}{\sqrt{3}}& \frac{4v_1^2+v_2^2+4v_3^2}{3}
\end{pmatrix}
\begin{pmatrix}
A^{3\mu} \\A^{6\mu} \\ A^{8\mu}
\end{pmatrix}\notag\\&+\frac{\tilde{g}^2}{8}(v_1^2+v_2^2+v_3^2)A^7_\mu A^{7\mu}.
\end{align}
We note that the mixing between $A^1_\mu-A^4_\mu$ and $A^2_\mu-A^5_\mu$ pairs
is the same and vanishes when $v_3=0$. Similarly, $A^6_\mu$ becomes an eigenstate degenerate 
with $A^7_\mu$ when $v_3=0$, however there is still mixing between $A^3_\mu$ and $A^8_\mu$.


\end{document}